\documentclass[12pt]{article}
\usepackage{amssymb,amsmath}
\makeatletter

\usepackage{arydshln}

\newcount\comment@nesting

\begingroup
\xdef\comment@begincomment{\string\\begin\string\{comment\string\}}
\xdef\comment@endcomment{\string\\end\string\{comment\string\}}
\endgroup

\begingroup
\catcode`\^^M=\active
\def\@temp{\endgroup\def\comment@processline##1^^M}%
\@temp{%
    \def\comment@curline{#1}%
    \let\@next=\comment@processline
    \ifx\comment@curline\comment@endcomment
        \ifnum\comment@nesting=0
            \def\@next{\end{comment}}%
        \else
            \global\advance\comment@nesting by -1
        \fi
    \else
        \ifx\comment@curline\comment@begincomment
            \global\advance\comment@nesting by 1
        \fi
    \fi
    \@next
}

\makeatother

\usepackage[dvipdfmx]{graphicx}
\usepackage{bm}
\usepackage{color}
\usepackage{here}
\usepackage{enumerate}
\usepackage{here}
\usepackage{subfigure}
\usepackage{tikz}
\usepackage[T1]{fontenc}
\usepackage{hyperref}
\newcommand{\Zb}{\mathbb{Z}}

\newcommand{\Acal}{\mathcal{A}}

\newcommand{\Dcal}{\mathcal{D}}

\newcommand{\Ncal}{\mathcal{N}}

\DeclareMathOperator*{\Tr}{{\rm Tr}}

\newcommand{\rII}{\mathcal{I} \hspace*{-0.15cm} \mathcal{I}}
\newcommand{\II}{\mathbb{II}}

\numberwithin{equation}{section}

\allowdisplaybreaks[1]

\definecolor{mygreen}{rgb}{0,0.714,0.286}

\begin{document}

%%%%%%%%%%%%%%%%%%%%%%%%%%%%%%%%%%%%%%%%%%%%
\thispagestyle{empty}
\begin{flushright}
%%%%%%%%%%%%%%%%%%%%%%%%%%%%%%%%%%%%%%%%%%%%%%%%%%%%%%%%%%%%%%%%%
%input \\
%%%%%%%%%%%%%%%%%%%%%%%%%%%%%%%%%%%%%%%%%%%%%%%%%%%%%%%%%%%%%%%%%

\end{flushright}
\vskip1.5cm
\begin{center}
{\Large \bf 
3d exceptional gauge theories
\\
\vskip0.75cm
and boundary confinement}

\vskip1.5cm
Tadashi Okazaki\footnote{tokazaki@seu.edu.cn}

\bigskip
{\it Shing-Tung Yau Center of Southeast University,\\
Yifu Architecture Building, No.2 Sipailou, Xuanwu district, \\
Nanjing, Jiangsu, 210096, China
}

\bigskip
and
\\
\bigskip
Douglas J. Smith\footnote{douglas.smith@durham.ac.uk}

\bigskip
{\it Department of Mathematical Sciences, Durham University,\\
Upper Mountjoy, Stockton Road, Durham DH1 3LE, UK}

\end{center}

%%%%%%%%%%%%%%%%%%%%%%%%%%%%%%%%%%%%%%%%%%%%
\vskip1cm
\begin{abstract}
We find boundary confining dualities of 3d $\mathcal{N}=2$ supersymmetric gauge theories with exceptional gauge groups.
The half-indices which enumerate the boundary BPS local operators in the presence of Neumann and Dirichlet boundary conditions for gauge fields are identified with 
the Askey-Wilson type $q$-beta integrals and Macdonald type sums respectively. 
New conjectural identities of $E_6$ and $E_7$ integrals and sums
are derived
from the boundary confining dualities. 
We also consider theories with a vector multiplet and adjoint chiral, which correspond to an $\mathcal{N}=4$ vector multiplet, with appropriate boundary conditions. We argue for the boundary confinement of the $\mathcal{N}=4$ vector multiplet and comment on such theories also with classical gauge groups.
\end{abstract}
%%%%%%%%%%%%%%%%%%%%%%%%%%%%%%%%%%%%%%%%%%

\newpage
\setcounter{tocdepth}{3}
\tableofcontents
%%%%%%%%%%%%%%%%%%%%%%%%%%

%%%%%%%%%%%%%%%%%%%%%%%%%%%%%%%%%%
%%%%%%%%%%%%%%%%%%%%%%%%%%%%%%%%%%
\section{Introduction}
\label{sec_Intro}
%%%%%%%%%%%%%%%%%%%%%%%%%%%%%%%%%%
%%%%%%%%%%%%%%%%%%%%%%%%%%%%%%%%%%
A confining duality allows for a dual description of a gauge theory 
as a non-gauge theory in terms of gauge invariant composites and their interactions. 
It can be used to construct a sequence of dual theories and to figure out an essential or basic part of dualities.
For supersymmetric non-Abelian gauge theories the confining dualities have been found for various theories 
e.g. for 4d $\mathcal{N}=1$ supersymmetric gauge theories 
\cite{Intriligator:1995ne,Berkooz:1995km,Pouliot:1995me,Luty:1996cg,Csaki:1996sm,Csaki:1996zb,Terning:1997jj,Garcia-Etxebarria:2012ypj,Garcia-Etxebarria:2013tba,Etxebarria:2021lmq,Bajeot:2022lah,Bajeot:2022kwt,Bottini:2022vpy,Amariti:2023wts}, 
3d $\mathcal{N}=2$ supersymmetric gauge theories 
\cite{Amariti:2015kha,Nii:2016jzi,Pasquetti:2019uop,Pasquetti:2019tix,Nii:2019dwi,Benvenuti:2020gvy,Benvenuti:2021nwt,Bajeot:2022lah,Amariti:2022wae,Okazaki:2023hiv,Amariti:2023wts} 
and 2d $\mathcal{N}=(0,2)$ supersymmetric gauge theories \cite{Sacchi:2020pet}. 
Recently in \cite{Okazaki:2023hiv} we found boundary confining dualities of 3d $\mathcal{N}=2$ supersymmetric gauge theories, 
whose vector multiplets of gauge groups $SU(N)$, $USp(2n)$ and $SO(N)$ obey the half-BPS $\mathcal{N}=(0,2)$ Neumann boundary conditions due to the presence of a boundary 
(See e.g. \cite{Gadde:2013wq,Okazaki:2013kaa,Gadde:2013sca,Yoshida:2014ssa,Dimofte:2017tpi,Brunner:2019qyf,Costello:2020ndc,Sugiyama:2020uqh,Okazaki:2021pnc,Okazaki:2021gkk,Alekseev:2022gnr,Dedushenko:2022fmc,Okazaki:2023hiv,Crew:2023tky,Dedushenko:2023qjq} for the study of the half-BPS $\mathcal{N}=(0,2)$ boundary conditions in 3d $\mathcal{N}=2$ supersymmetric gauge theories). 
Similarly to the theories in the bulk, the dual boundary descriptions are given by non-gauge theories obeying certain $\mathcal{N}=(0,2)$ boundary conditions. 
The boundary confining dualities result in the equivalences of the half-indices \cite{Gadde:2013wq,Gadde:2013sca,Yoshida:2014ssa,Dimofte:2017tpi} which count the BPS local operators living at the boundary with appropriate boundary conditions. 
They are identified with the Askey-Wilson type $q$-beta integrals 
\cite{MR783216,MR772878,MR845667,MR1139492,MR1266569,MR2267266} 
associated with the root systems corresponding to the gauge groups so that the boundary confining dualities can be rigorously confirmed as identities of the matrix integrals. 
Moreover, one can find new integral formulas from the boundary confining dualities for 3d $\mathcal{N}=2$ gauge theories and vice versa. 

In this paper, we describe 3d $\mathcal{N} = 2$ boundary confining dualities where the gauge group is an exceptional group.
While supersymmetric gauge theories with exceptional gauge groups are examined in the literature, e.g.\ \cite{Ramond:1996ku,Distler:1996ub,Karch:1997jp,Cho:1997am,Grinstein:1998bu,Pouliot:2001iw} for 4d $\mathcal{N}=1$ theories, in \cite{Nii:2017npz,Nii:2019dwi,Nii:2019wjz} for 3d $\mathcal{N}=2$ theories and in \cite{Chen:2018wep} for 2d $\mathcal{N}=(2,2)$ theories, 
the dualities are less well understood.  
With the tool of half-indices and boundary 't Hooft anomalies we find strong evidence of boundary confining dualities for exceptional gauge theories. 
In one type the theory consists of the vector multiplet and an adjoint chiral, both with Neumann boundary conditions. Such theories provide examples of boundary confining theories for any gauge group. The matching of half-indices corresponds to identities known as the orthogonality measures for the Macdonald polynomials \cite{MR674768,MR1314036}. 
The other examples we consider have fundamental chirals (or both fundamental and antifundamental for gauge group $E_6$), again with Neumann boundary conditions for all chirals. In the case of gauge group $G_2$ or $F_4$ the matching half-indices beautifully reproduces known identities. For $E_6$ or $E_7$ 
the integral formula is unknown in the literature. 
So we obtain new conjectured identities from the boundary confining dualities.
%these are new conjectured identities.
We can also exchange all Neumann and Dirichlet boundary conditions. The theory A half-indices then correspond to Macdonald type sums \cite{MR674768, MR2018362}. The theory B half-indices then give the evaluation of these sums, corresponding to Macdonald type identities.
See e.g.\ \cite{MR2267266} with proofs in \cite{MR1868358, MR1926355, MR2016669} for a summary of known cases which relate to boundary confining dualities. Our examples of $E_6$ or $E_7$ gauge group with fundamentals provide new conjectured identities for generalized Macdonald type sums.

%%%%%%%%%%%%%%%%%%%%%%%%%%%%%%%%%%
\subsection{Open problems}
\label{sec_openQ}
%%%%%%%%%%%%%%%%%%%%%%%%%%%%%%%%%%
We list several open problems which we leave for future works.
\begin{itemize}
\item 
%more general exceptional dualities 
Unlike the case with unitary, orthogonal and symplectic gauge groups, 
the dualities of gauge theories with exceptional gauge theories are less understood. 
It would be intriguing to explore such dualities including the Chern-Simons coupling 
\footnote{
See \cite{Cordova:2018qvg} for dualities of exceptional Chern-Simons theories.
}

\item 
%analytic proof
While we have provided several numerical and physical checks for the conjectural identities of the half-indices 
as the unknown evaluations of $E_6$ and $E_7$ Askey-Wilson type $q$-beta integrals, 
it would be nice to give an analytic proof with some extra work as discussed for other root systems \cite{MR783216,MR772878,MR845667,MR1139492,MR1266569,MR2267266}. 
\item 
%line operators
The half-index can be generaliezd by line defect operators ending on the boundary. 
When the Wilson line operator transforming in the representation $\mathcal{R}$ is introduced, the integrand of the Askey-Wilson type $q$-beta integral is decorated 
by the characters of $\mathcal{R}$. It would be interesting to explore dualities with such line operators as well as understand the mathematical relation of such half-indices to Askey-Wilson polynomial orthogonality relations.
\item 
%line operators
The Neumann half-index for the vector multiplet can be enriched by 
introducing 2d degrees of freedom at the boundary. 
The generalized confinement/duality of the 3d-2d coupled system will be addressed by gauging the Dirichlet half-indices as employed in \cite{Okazaki:2019bok}.
%recent confining duality (Benvenuti etal) and their boundary confining duality
\end{itemize}

%%%%%%%%%%%%%%%%%%%%%%%%%%%%%%%%%%
\subsection{Structure}
\label{sec_str}
The structure of the paper is straightforward. 
In section \ref{sec_RedHalfInd} we review the $\mathcal{N}=(0,2)$ half-indices of 3d $\mathcal{N}=2$ theories 
and introduce the ``reduced half-indices'', which are obtained by taking a limit of fugacities. 
In section \ref{sec_G_adj} we discuss the basic features which are required 
to calculate the boundary 't Hooft anomalies and the half-indices of the gauge theories with an adjoint chiral. 
In section \ref{sec_G2} we find the $\mathcal{N}=(0,2)$ boundary confining dualities for $G_2$ gauge theory with 4 fundamental chirals 
and that with an adjoint chiral. 
In section \ref{sec_F4} we present the $\mathcal{N}=(0,2)$ boundary confining dualities for $F_4$ gauge theories with 3 fundamental chirals 
and that with an adjoint chiral. 
In section \ref{sec_E6} we find the $\mathcal{N}=(0,2)$ boundary confining dualities for $E_6$ gauge theories with $(N_f,N_a)$ 
$=$ $(4,0)$, $(3,1)$ and $(2,2)$ where $N_f$ (resp. $N_a$) is the number of the fundamental (resp. antifundamental) chirals 
as well as that with an adjoint chiral. 
In section \ref{sec_E7} the $\mathcal{N}=(0,2)$ boundary confining dualities for $E_7$ gauge theories with 3 fundamental chirals and that with an ajoint chiral 
are presented. 
In section \ref{sec_E8_adj_integral} we propose the $\mathcal{N}=(0,2)$ boundary confining duality for $E_8$ gauge theory with an adjoint chiral. 
%%%%%%%%%%%%%%%%%%%%%%%%%%%%%%%%%%

%%%%%%%%%%%%%%%%%%%%%%%%%%%%%%%%%%
%%%%%%%%%%%%%%%%%%%%%%%%%%%%%%%%%%
\section{Half-indices and reduced half-indices}
\label{sec_RedHalfInd}
%%%%%%%%%%%%%%%%%%%%%%%%%%%%%%%%%%
%%%%%%%%%%%%%%%%%%%%%%%%%%%%%%%%%%
Analogously to the definition of 3d superconformal indices, the supersymmetric ``half-index'' of a 3d $\mathcal{N}=2$ supersymmetric theory $\mathcal{T}$ obeying the $\mathcal{N}=(0,2)$ half-BPS boundary condition $\mathcal{B}$
is defined as a trace over the states that correspond to half-BPS local operators on the boundary
\cite{Gadde:2013wq, Gadde:2013sca, Yoshida:2014ssa, Dimofte:2017tpi}
\begin{align}
\mathbb{II}_{\mathcal{B}}^{\mathcal{T}}&={\Tr}_{\mathrm{Op}} (-1)^F q^{J+\frac{R}{2}} a^{q_a} x^{f}
\end{align}
where $F$ is the Fermion number operator,
$J$ is the generator of the $Spin(2)\cong U(1)_J$ rotational symmetry of the two-dimensional plane where the boundary local operators are supported,
$R$ is the $U(1)_R$ R-charge, $q_a$ is the $U(1)_a$ axial charge (or we may have more than one axial charge)
and $f$ the Cartan generators of the remaining global symmetry group which will be a (product of) special unitary group(s) in our examples.

The half-index can be computed as a partition function on $HS^2\times S^1$
which encodes the $\mathcal{N}=(0,2)$ half-BPS boundary conditions on $\partial(HS^2\times S^1)=S^1\times S^1$
where $HS^2$ is a hemisphere.
The UV formula for the half-index of the Neumann boundary condition for the gauge multiplet was derived in \cite{Gadde:2013wq, Gadde:2013sca, Yoshida:2014ssa}
and the formula for the Dirichlet boundary conditions for the gauge multiplet was proposed in \cite{Dimofte:2017tpi}.
%The half-index can also realize the holomorphic block \cite{Pasquetti:2011fj, Beem:2012mb},
%the $q$-series 3-manifold invariant $\hat{Z}$ \cite{Gukov:2016gkn} which has a number of applications to topology and number theory
%(see e.g. \cite{Gukov:2017kmk, Cheng:2018vpl, Gukov:2019mnk, Chun:2019mal, Chung:2019khu, Ekholm:2020lqy})
%and the 4-manifold invariant \cite{Vafa:1994tf} (see e.g. \cite{Feigin:2018bkf}).

We will give explicit details of the half-index expressions for the cases relevant to the theories discussed in this article. However, a common feature is that the half-indices can be expressed in terms of $q$-Pochhammer symbols $(X; q)_{\infty} = \prod_{n = 0}^{\infty} (1 - Xq^n)$ where $X$ may contain a fixed power of $q$. In cases where a vector multiplet has Neumann boundary conditions we will have an integral over the gauge fugacities while for Dirichlet boundary conditions we will instead have a sum over monopole fluxes. For cases where there is no vector multiplet we have simply a rational function of these $q$-Pochhammer symbols. There is an obvious simplification by taking a limit $q \to 0$ which results in simply $(1-X)$ if $X$ is fixed in this limit (or even more simply $1$ if $X \to 0$ in this limit). As we will see, such limits are well-defined in the examples we consider where we have a vector multiplet with Neumann boundary condition and in the duals of these boundary confining gauge theories. In such cases the expressions $X$ will be combinations of the global symmetry fugacities such as global $U(1)$ (e.g. axial symmetry) fugacities and non-Abelian flavor fugacities. For example, we typically shift the $U(1)_R$ charge by an amount proportional to the $U(1)_a$ charge which corresponds to replacing the axial fugacity $a \to q^{r/2}a$ but in the limit $q \to 0$ we keep $\mathfrak{t} = q^{r/2}a$ fixed. We will refer to such limits of half-indices as reduced half-indices, $\rII$. Specifically, with the understanding that certain combinations of fugacities are fixed,
\begin{align}
\rII_{\mathcal{B}}^{\mathcal{T}} (\mathfrak{t}, x) & = \lim_{q \to 0} \mathbb{II}_{\mathcal{B}}^{\mathcal{T}} (\mathfrak{t}, x; q) \; .
\end{align}
These reduced half-indices have the form of 
Hilbert series or graded trace for the algebra formed by the boundary local operators
%boundary versions of Hilbert series
if we also set the non-Abelian flavor symmetry fugacities to $1$ or of refined Hilbert series if we keep them.

In cases where we do not have a proof of the matching of half-indices or even the resource to calculate low order terms in the $q$-series expansion of half-indices for high rank gauge groups (specifically $E_6$ and $E_7$ gauge theories with Neumann boundary conditions for the vector multiplet) we can still provide some evidence for dualities by showing that the reduced half-indices match.

%%%%%%%%%%%%%%%%%%%%%%%%%%%%%%%%%%
%%%%%%%%%%%%%%%%%%%%%%%%%%%%%%%%%%
\section{Gauge theory with an adjoint chiral}
\label{sec_G_adj}
%%%%%%%%%%%%%%%%%%%%%%%%%%%%%%%%%%
%%%%%%%%%%%%%%%%%%%%%%%%%%%%%%%%%%
We will focus on $\mathcal{N} = 2$ boundary confining theories with exceptional gauge groups. These will include cases where there is an adjoint chiral in addition to the vector multiplet which can also be viewed as $\mathcal{N} = 4$ boundary confining theories with only a vector multiplet. Here we will first comment on some general features of a gauge theory with gauge group $G$ of dimension $d_G$ and a single adjoint chiral.
If we take the same boundary conditions for the vector multiplet and the adjoint chiral, the gauge anomaly will cancel for any gauge group without introducing any other chirals or 2d matter. This corresponds to an $\mathcal{N} = (2,2)$ boundary condition for the $\mathcal{N} = 4$ theory and gives a candidate boundary confining theory A. We will later present explicit examples for exceptional gauge groups.

If we take the adjoint chiral to have zero R-charge and a $U(1)_a$ flavor symmetry charge $+1$ then with Neumann boundary conditions for both the vector multiplet and the adjoint chiral we have the following anomaly
\begin{align}
\label{bdy_G_adj_anomA}
\Acal & = \underbrace{{h^{\vee}_G \Tr(s^2)} + \frac{d_G}{2} r^2}_{\textrm{VM}, \; \Ncal}
 - \underbrace{\left( h^{\vee}_G \Tr(s^2) + \frac{d_G}{2} (a-r)^2 \right)}_{\Phi, \; N}
  \nonumber \\
  = & - \frac{d_G}{2} a^2 + d_G ar
\end{align}
where $h^{\vee}_G$ is the dual Coxeter number.

Assuming we have a boundary confining duality, the dual theory will have chirals $M_{d_f}$ with Neumann boundary conditions with $U(1)_a$ charges $d_f$ where $d_f$ are the degrees of the independent invariants which can be formed from an adjoint scalar. These are the same as the degrees of the fundamental invariants of the Coxeter group corresponding to the Weyl group of $G$. These degrees are
\begin{align}
\label{G_fund_inv}
\begin{array}{c|c|c}
G & d_G & d_f \\
\hline
SU(N) & N^2 - 1 & 2, 3, 4, \ldots N \\
USp(2N) & N(2N+1) & 2, 4, 6, \ldots 2N \\
SO(2N) & N (2N-1)& N; 2, 4, 6, \ldots 2N-2 \\
SO(2N+1) & N(2N+1) & 2, 4, 6, \ldots 2N \\
G_2 & 14 & 2, 6 \\
F_4 & 52 & 2, 6, 8, 12 \\
E_6 & 78 & 2, 5, 6, 8, 9, 12 \\
E_7 & 133 & 2, 6, 8, 10, 12, 14, 18 \\
E_8 & 248 & 2, 8, 12, 14, 18, 20, 24, 30
\end{array}
\end{align}
It is straightforward to check (for any special unitary, orthogonal, symplectic or exceptional group $G$) that the anomalies match if we pair each of these chirals $M_{d_f}$ with a chiral $V_{d_f - 1}$ with Dirichlet boundary condition having $U(1)_a$ charge $1 - d_f$, and we take all chirals to have R-charge zero. The matching of anomalies is then equivalent to the identity
\begin{align}
    \label{df_dG}
    \sum_{d_f} \left( 2d_f - 1 \right) & = d_G
\end{align}
which is satisfied for all cases in (\ref{G_fund_inv}). In all these cases there is no standard unitary bulk duality since all the chirals have R-charge $0$ but since some have positive axial charge and others negative axial charge, we cannot shift the R-charge by an amount proportional to the axial charge without leaving some chirals with negative dimension. Indeed, these are well known to be `bad' \cite{Gaiotto:2008ak} $\mathcal{N} = 4$ theories. What we are claiming here is that these theories are well defined with Neumann boundary conditions and that they are then boundary confining theories with duals described above. It is also possible to switch all Neumann and Dirichlet boundary conditions to get another boundary confining duality for each group $G$.

While the existence of the chirals $M_{d_f}$ in theory B, corresponding to invariants of the adjoint scalar in theory A, is not surprising, the appearance of the chirals $V_{d_f - 1}$ is less obvious. However, these are precisely what is required in order to realise half-BPS boundary conditions for 3d $\mathcal{N} = 4$ gauge theories. Indeed, in theory A the $\mathcal{N} = 2$ vector multiplet and adjoint chiral multiplet with Neumann boundary conditions correspond to an $\mathcal{N} = 4$ vector multiplet with Neumann boundary conditions. In terms of the half-index the correspondence is
\begin{align}
    \frac{(s; q)_{\infty}}{(q^{r/2} a s; q)_{\infty}} & = \frac{(s; q)_{\infty}}{(q^{1/2} t^{-2} s; q)_{\infty}}
\end{align}
where there will be a product over $s$ which represents some combination of gauge fugacities, $a$ is the $U(1)_a$ fugacity and $t$ is the $U(1)_{H-C} \subset SU(2)_C \times SU(2)_H$ fugacity, following the conventions in \cite{Okazaki:2020lfy}. On the left hand side, the numerator is the contribution from the $\mathcal{N} = 2$ vector multiplet while the denominator is the contribution from the adjoint chiral.
Similarly, in theory B the pair $M_{d_f}$ with Neumann boundary conditions and $V_{d_f - 1}$ with Dirichlet boundary conditions corresponds to an $\mathcal{N} = 4$ twisted hypermultiplet. In terms of the half-index the correspondence is
\begin{align}
    \frac{(q^{1 + (d_f - 1)r/2} a^{d_f - 1}; q)_{\infty}}{(q^{d_f r/2} a^{d_f}; q)_{\infty}} & = \frac{(q^{3/4} t x; q)_{\infty}}{(q^{1/4} t^{-1} x; q)_{\infty}}
\end{align}
where $x$ is the global fugacity for the $\mathcal{N} = 4$ twisted hypermultiplet. It is straightforward to see that the $\mathcal{N} = 2$ and $\mathcal{N} = 4$ descriptions match with the identifications
\begin{align}
    t & = q^{(1-r)/4} a^{-1/2} \\
    x & = q^{d_f r/2 - r/4} a^{d_f - 1/2}
\end{align}
which identifies $U(1)_a$ as a $U(1)$ subgroup of the $\mathcal{N} = 4$ $R$-symmetry. We then see that the $U(1)_x$ global symmetry is broken with $U(1)_x$ transformations compensated by specific $U(1)_R$ and $U(1)_a$ transformations in the $\mathcal{N} = 2$ description. We interpret this a specific type of $\mathcal{N} = 4$ $D_c$ boundary condition.

Note that for $G = SU(N)$ the fundamental invariants are $\Tr \Phi^{d_f}$ where $d_f$ can take the values $2, 3, \ldots, N$. Such a boundary duality for $G = SU(2)$ has been described in \cite{Dimofte:2017tpi} where indeed the dual theory consists of a chiral with Neumann boundary condition and $U(1)_a$ charge $2$, and a chiral with Dirichlet boundary condition and $U(1)_a$ charge $-1$.

More generally, for Neumann boundary conditions in theory A the matching of half-indices for any group $G$ is equivalent to the identity conjectured by Macdonald \cite{MR674768} 
and later proven by Cherednik \cite{MR1314036}. 
\footnote{
Although we will not pursue it here, this can be also viewed as the identity of 4d half-indices, which implies the S-duality of Neumann boundary conditions and Nahm pole boundary conditions in 4d $\mathcal{N}=4$ super Yang-Mills theory as demonstrated in \cite{Gaiotto:2019jvo}. }
In later sections we will present the explicit results for exceptional gauge groups.

%%%%%%%%%%%%%%%%%%%%%%%%%%%%%%%%%%
%%%%%%%%%%%%%%%%%%%%%%%%%%%%%%%%%%
\section{Gauge group $G_2$}
\label{sec_G2}
%%%%%%%%%%%%%%%%%%%%%%%%%%%%%%%%%%
%%%%%%%%%%%%%%%%%%%%%%%%%%%%%%%%%%
In this section we consider boundary confining dualities for $G_2$ gauge theories with fundamental or adjoint chirals. We start with the theory with $4$ fundamental chirals then consider instead an adjoint chiral.

%%%%%%%%%%%%%%%%%%%%%%%%%%%%%%%%%%
%%%%%%%%%%%%%%%%%%%%%%%%%%%%%%%%%%
\subsection{$G_2$ with $4$ fundamental chirals}
\label{sec_G2_4_integral}
%%%%%%%%%%%%%%%%%%%%%%%%%%%%%%%%%%
%%%%%%%%%%%%%%%%%%%%%%%%%%%%%%%%%%
Consider 3d $\mathcal{N}=2$ gauge theory with gauge group $G_2$ and $4$ fundamental chirals $Q_{\alpha}$, $\alpha=1,\cdots, 4$. 
It is argued in \cite{Nii:2017npz} that the theory is confining.  
\footnote{See \cite{Pesando:1995bq,Giddings:1995ns} for the study of 4d $\mathcal{N}=1$ $G_2$ gauge theory. }
We now consider the case with boundary conditions and show that there is a set of boundary conditions for which the gauge anomaly is cancelled in the $G_2$ gauge theory and that the 't Hooft anomalies match for the gauge theory and its dual. We present the matching half-indices, noting that this is equivalent to an identity proven by Gustafson \cite{MR1139492}.

We consider first the case where the vector and chiral multiplets obey Neumann boundary conditions. We then propose a
dual theory B with the field content (the same as in the bulk dual discussed in \cite{Nii:2017npz}) and boundary conditions:
an $SU(4)$ rank-$2$ symmetric chiral $M_{\alpha\beta}$, an $SU(4)$ antifundamental chiral $B^{\alpha}$, a singlet $B$ all obeying Neumann boundary conditions, 
as well as a singlet $V$ with Dirichlet boundary condition.  
The boundary conditions and charges of the field content are summarized as follows: 
\begin{align}
\label{G2_4_charges}
\begin{array}{c|c|c|c|c|c}
& \textrm{bc} & G_2 & SU(N_f = 4) & U(1)_a & U(1)_R \\ \hline
\textrm{VM} & \mathcal{N} & {\bf Adj} & {\bf 1} & 0 & 0 \\
Q_{\alpha} & \textrm{N} & {\bf 7} & {\bf 4} & 1 & 0 \\
 \hline
M_{\alpha \beta} & \textrm{N} & {\bf 1} & {\bf 10} & 2 & 0 \\
B^{\alpha} & \textrm{N} & {\bf 1} & {\bf \overline{4}} & 3 & 0 \\
B & \textrm{N} & {\bf 1} & {\bf 1} & 4 & 0 \\
V & \textrm{D} & {\bf 1} & {\bf 1} & -8 & 2
\end{array}
\end{align}
The operators map as $M_{\alpha \beta} \sim Q_{\alpha}Q_{\beta}$ with the gauge indices contracted with the symmetric rank-$2$ $G_2$-invariant tensor, $B^{\alpha} \sim Q_{\beta_1}Q_{\beta_2}Q_{\beta_3} \epsilon^{\alpha \beta_1 \beta_2 \beta_3}$ with the gauge indices contracted with the antisymmetric rank-$3$ $G_2$-invariant tensor, $B \sim Q_1 Q_2 Q_3 Q_4$ with the gauge indices contracted with the antisymmetric rank-$4$ $G_2$-invariant tensor, and $V$ (in the bulk theory) is dual to the minimal monopole in theory A.

We can easily check there is no gauge anomaly and the 't Hooft anomalies match as follows,
\begin{align}
\label{bdy_G2_4_anomAB}
\Acal & = \underbrace{{4\Tr(s^2)} + 7r^2}_{\textrm{VM}, \; \Ncal}
 - \underbrace{\left( 4\Tr(s^2) + \frac{7}{2}\Tr(x^2) +14(a-r)^2 \right)}_{Q_{\alpha}, \; N}
  \nonumber \\
  = & - \frac{7}{2} \Tr(x^2) - 14a^2 + 28ar - 7r^2
  \nonumber \\
  & = - \underbrace{\left( 3\Tr(x^2) + 5(2a-r)^2 \right)}_{M_{\alpha \beta}, \; N}
   + \underbrace{\frac{1}{2}\big( -8a + r \big)^2}_{V, \; D}
  \nonumber \\
  & - \underbrace{\left( \frac{1}{2}\Tr(x^2) + 2(3a-r)^2 \right)}_{B^{\alpha}, \; N}
   - \underbrace{\frac{1}{2}\big( 4a - r \big)^2}_{B, \; N} \; .
\end{align}

The half-index for theory A is
\begin{align}
\label{bdy_G2_4_hindexA}
\II^A_{\Ncal, N} = & \frac{(q)_{\infty}^2}{2^2 3} \prod_{i=1}^2 \oint \frac{ds_i}{2\pi i s_i}
\frac{\left( \prod_{i \ne j}^3 (s_i s_j^{-1}; q)_{\infty} \right) \prod_{i = 1}^3 (s_i^{\pm}; q)_{\infty} }{\prod_{\alpha = 1}^4 (q^{r/2} a x_{\alpha}; q)_{\infty} \prod_{i = 1}^3 (q^{r/2} s_i^{\pm} a x_{\alpha}; q)_{\infty}}, 
\end{align}
where $(X^{\pm};q)_{\infty}$ $=$ $(X;q)_{\infty}(X^{-1};q)_{\infty}$ and
$\prod_{i = 1}^3 s_i = \prod_{\alpha = 1}^4 x_{\alpha} = 1$. 
When we set $r=1/4$ and $x_\alpha=1$, we get
\begin{align}
1+10 a^2 q^{1/4}+4 a^3 q^{3/8}+56 a^4q^{1/2}+40 a^5 q^{5/8}+240 a^6 q^{3/4}+224 a^7 q^{7/8}+\cdots. 
\end{align}

According to the identity proven by Gustafson \cite{MR1139492}, 
it follows that the integral (\ref{bdy_G2_4_hindexA}) is equal to 
\begin{align}
\label{bdy_G2_4_hindexB}
\II^B_{N, N, N, D} = & \frac{(q^{4r}a^8;q)_{\infty}}{(q^{2r}a^4;q)_{\infty} \left( \prod_{\alpha \le \beta}^4 (q^r a^2 x_{\alpha} x_{\beta}; q)_{\infty} \right) \prod_{\alpha = 1}^4 (q^{3r/2} a^3 x_{\alpha}^{-1};q)_{\infty}}. 
\end{align}
The expression (\ref{bdy_G2_4_hindexB}) is precisely the half-index of dual theory B with the above boundary conditions.

We can also take a suitable limit with $q \to 0$ to produce a reduced half-index. In this case for theory A we first fix $\mathfrak{t} = q^{r/2}a$ and $x_{\alpha}$ which gives
\begin{align}
\label{bdy_G2_4_HseriesA}
\rII^A_{\Ncal, N} = & \frac{1}{2^2 3} \prod_{i=1}^2 \oint \frac{ds_i}{2\pi i s_i}
\frac{\left( \prod_{i \ne j}^3 (1 - s_i s_j^{-1}) \right) \prod_{i = 1}^3 (1 - s_i^{\pm}) }{\prod_{\alpha = 1}^4 (1 - \mathfrak{t} x_{\alpha}) \prod_{i = 1}^3 (1 - s_i^{\pm} \mathfrak{t} x_{\alpha})}
\end{align}
The integrand has simple poles at
\begin{align}
(s_1, s_2)  = (\mathfrak{t} x_{\beta}, \mathfrak{t} x_{\gamma}) \; , \;\; \beta \ne \gamma \\
(s_1, s_2)  = (\mathfrak{t} x_{\beta}, x_{\beta}^{-1} x_{\gamma}) \; , \;\; \beta > \gamma \\
(s_1, s_2)  = (s_2^{-1} \mathfrak{t} x_{\beta}, \mathfrak{t} x_{\gamma}) \; , \;\; \beta \ne \gamma \\
(s_1, s_2)  = (s_2^{-1} \mathfrak{t} x_{\beta}, \mathfrak{t}^2 x_{\beta} x_{\gamma}) \; , \;\; \beta \ne \gamma \\
(s_1, s_2)  = (s_2^{-1} \mathfrak{t} x_{\beta}, x_{\beta} x_{\gamma}^{-1}) \; , \;\; \beta < \gamma
\end{align}
when integrating first wrt.\ $s_1$ and ordering $|x_{\beta}| < |x_{\gamma}|$ for $\beta < \gamma$ in the case that $|\mathfrak{t}|$ is sufficiently small so that $|\mathfrak{t} x_4| < 1$. It is straightforward to evaluate the integral by summing the residues of these poles. The result is precisely the reduced half-index for theory B
\begin{align}
\label{bdy_G2_4_HseriesB}
\rII^B_{N, N, N, D} = & \frac{(1 + \mathfrak{t}^4)}{\left( \prod_{\alpha \le \beta}^4 (1 - \mathfrak{t}^2 x_{\alpha} x_{\beta}) \right) \prod_{\alpha = 1}^4 (1 - \mathfrak{t}^3 x_{\alpha}^{-1})}. 
\end{align}

We can also exchange all Neumann and Dirichlet boundary conditions which just switches the sign of all contributions to the anomalies so the anomalies will still match. Doing so gives the theory A half-index
\begin{align}
\label{bdy_G2_4_hindexA_D}
\II^A_{\Dcal, D} = & \frac{1}{(q)_{\infty}^2} \sum_{m_1, m_2 \in \Zb} \frac{\prod_{\alpha = 1}^4 (q^{1 - r/2} a^{-1} x_{\alpha}^{-1}; q)_{\infty} \prod_{i = 1}^3 (q^{1 - r/2 \pm m_i} s_i^{\pm} a^{-1} x_{\alpha}^{-1}; q)_{\infty}}{\left( \prod_{i \ne j}^3 (q^{1 + m_i - m_j}s_i s_j^{-1}; q)_{\infty} \right) \prod_{i = 1}^3 (q^{1 \pm m_i}s_i^{\pm}; q)_{\infty}}, 
\end{align}
where $\prod_{i = 1}^3 s_i = \prod_{\alpha = 1}^4 x_{\alpha} = 1$ and $\sum_{i = 1}^3 m_i = 0$. The
theory B half-index is
\begin{align}
\label{bdy_G2_4_hindexB_D}
\II^B_{D, D, D, N} = & \frac{(q^{1 - 2r}a^{-4};q)_{\infty} \left( \prod_{\alpha \le \beta}^4 (q^{1 - r} a^{-2} x_{\alpha}^{-1} x_{\beta}^{-1}; q)_{\infty} \right) \prod_{\alpha = 1}^4 (q^{1 - 3r/2} a^{-3} x_{\alpha};q)_{\infty}}{(q^{1 - 4r}a^{-8};q)_{\infty}}. 
\end{align}

Note that the Dirichlet half-index for theory A is independent of the gauge fugacities which is not at all obvious from the expression (\ref{bdy_G2_4_hindexA_D}). Physically this is related to the confining duality. Mathematically the cases where this happens in such Macdonald type sums \cite{MR2018362} are listed, including the evaluation of the sums, in \cite{MR2267266} with proofs in \cite{MR1868358, MR1926355, MR2016669}. Indeed the above identity of half-indices is equivalent to a known evaluation of a Macdonald type sum.

%%%%%%%%%%%%%%%%%%%%%%%%%%%%%%%%%%
%%%%%%%%%%%%%%%%%%%%%%%%%%%%%%%%%%
\subsection{$G_2$ with an adjoint chiral}
\label{sec_G2_adj_integral}
%%%%%%%%%%%%%%%%%%%%%%%%%%%%%%%%%%
%%%%%%%%%%%%%%%%%%%%%%%%%%%%%%%%%%
For gauge group $G_2$ we consider Neumann boundary conditions for both the vector multiplet and an adjoint chiral. The charges and boundary conditions are summarized as follows: 
\begin{align}
\label{G2_adj_charges}
\begin{array}{c|c|c|c|c}
& \textrm{bc} & G_2 & U(1)_a & U(1)_R \\ \hline
\textrm{VM} & \mathcal{N} & {\bf Adj} & 0 & 0 \\
\Phi & \textrm{N} & {\bf Adj} & 1 & 0 \\
 \hline
M_2 & \textrm{N} & {\bf 1} & 2 & 0 \\
M_6 & \textrm{N} & {\bf 1} & 6 & 0 \\
V_1 & \textrm{D} & {\bf 1} & -1 & 0 \\
V_5 & \textrm{D} & {\bf 1} & -5 & 0
\end{array}
\end{align}
The boundary 't Hooft anomaly of theory A is given by
\begin{align}
\label{bdy_G2_adj_anomA}
\Acal ^A& = \underbrace{{4\Tr(s^2)} + 7r^2}_{\textrm{VM}, \; \Ncal}
 - \underbrace{\left( 4\Tr(s^2) +7(a-r)^2 \right)}_{\Phi, \; N}
  \nonumber \\
  = & - 7a^2 + 14ar , 
\end{align}
which exactly agrees with the boundary anomaly of theory B which is given by
\begin{align}
\label{bdy_G2_adj_anomB}
\Acal ^B= & - \underbrace{\frac{1}{2}(2a-r)^2}_{M_2, \; N}
   + \underbrace{\frac{1}{2}( -a - r )^2}_{V_1, \; D}
   - \underbrace{\frac{1}{2}(6a-r)^2}_{M_6, \; N}
   + \underbrace{\frac{1}{2}( -5a - r )^2}_{V_5, \; D} . 
\end{align}

This gives the half-index for theory A
\begin{align}
\label{bdy_G2_adj_hindexA}
\mathbb{II}^A_{\mathcal{N},N}
&=\frac{1}{2^2 3}
\frac{(q)_{\infty}^2}{(q^{r/2}a;q)_{\infty}^2}
\prod_{i=1}^{2}
\oint \frac{ds_i}{2\pi is_i}
\frac
{\left( \prod_{i \ne j}^3 (s_i s_j^{-1}; q)_{\infty} \right) \prod_{i = 1}^3 (s_i^{\pm}; q)_{\infty} }
{\left( \prod_{i \ne j}^3 (q^{r/2} a s_i s_j^{-1}; q)_{\infty} \right) \prod_{i = 1}^3 (q^{r/2} a s_i^{\pm}; q)_{\infty} }. 
\end{align}
For example, when $r=1/2$, we find
\begin{align}
\mathbb{II}^A_{\mathcal{N},N}&=
1+a^2 q^{1/2}+a^4 q-a q^{5/4}+\left(2 a^6+a^2\right) q^{3/2}-a^3 q^{7/4}+\left(2a^8+a^4\right) q^2
\nonumber\\
&+\left(-2 a^5-a\right) q^{9/4}+\left(2 a^{10}+2 a^6+a^2\right)
   q^{5/2}+\left(-3 a^7-2 a^3\right) q^{11/4}+\cdots. 
\end{align}

The half-index (\ref{bdy_G2_adj_hindexA}) coincides with the half-index for theory B
\begin{align}
\label{bdy_G2_adj_hindexB}
\mathbb{II}^B_{N, N, D, D} & =
\frac{
(q^{1+r/2}a;q)_{\infty}
(q^{1+5r/2}a^5;q)_{\infty}
}
{
(q^{r}a^2;q)_{\infty}
(q^{3r}a^6;q)_{\infty}
}, 
\end{align}
as shown by Cherednik \cite{MR1314036}. 

We can also exchange all Neumann and Dirichlet boundary conditions which just switches the sign of the contributions to the anomaly so the anomalies will still match. Doing so gives the theory A half-index
\begin{align}
\label{bdy_G2_adj_hindexA_D}
\II^A_{\Dcal, D} = & \frac{(q^{1 - r/2}a^{-1}; q)_{\infty}^2}{(q)_{\infty}^2} \sum_{m_1, m_2 \in \Zb} \frac{\left( \prod_{i \ne j}^3 (q^{1 - r/2 + m_i - m_j}a^{-1} s_i s_j^{-1}; q)_{\infty} \right) \prod_{i = 1}^3 (q^{1 - r/2 \pm m_i}a^{-1} s_i^{\pm}; q)_{\infty}}{\left( \prod_{i \ne j}^3 (q^{1 + m_i - m_j}s_i s_j^{-1}; q)_{\infty} \right) \prod_{i = 1}^3 (q^{1 \pm m_i}s_i^{\pm}; q)_{\infty}}, 
\end{align}
where $\prod_{i = 1}^3 s_i = 1$ and $\sum_{i = 1}^3 m_i = 0$. The
theory B half-index is
\begin{align}
\label{bdy_G2_adj_hindexB_D}
\II^B_{D, D, N, N} = & \frac{(q^{1 - r}a^{-2};q)_{\infty} (q^{1 - 3r} a^{-6}; q)_{\infty}}{(q^{- r/2}a^{-1};q)_{\infty} (q^{- 5r/2} a^{-5}; q)_{\infty}},
\end{align}
as proven in \cite{MR1868358, MR2016669}.

Note that the full indices won't be convergent since for $r \ne 0$ theory B will contain chirals of both positive and negative dimension. Similarly, in theory A if $r>0$ so that $\Phi$ has positive dimension then there will be negative dimension monopoles.

%%%%%%%%%%%%%%%%%%%%%%%%%%%%%%%%%%
%%%%%%%%%%%%%%%%%%%%%%%%%%%%%%%%%%
\section{Gauge group $F_4$}
\label{sec_F4}
%%%%%%%%%%%%%%%%%%%%%%%%%%%%%%%%%%
%%%%%%%%%%%%%%%%%%%%%%%%%%%%%%%%%%
In this section we consider boundary confining dualities for $F_4$ gauge theories with fundamental or adjoint chirals.

%%%%%%%%%%%%%%%%%%%%%%%%%%%%%%%%%%
%%%%%%%%%%%%%%%%%%%%%%%%%%%%%%%%%%
\subsection{$F_4$ with $3$ fundamental chirals}
\label{sec_F4_3_integral}
%%%%%%%%%%%%%%%%%%%%%%%%%%%%%%%%%%
%%%%%%%%%%%%%%%%%%%%%%%%%%%%%%%%%%
For gauge group $F_4$ with $3$ fundamental chirals there is a 3d bulk s-confinement described in \cite{Nii:2019dwi}. We propose a boundary confining duality by taking Neumann boundary conditions for all chirals except one in the dual theory. The content and boundary conditions are given by
\begin{align}
\label{F4_3_charges}
\begin{array}{c|c|c|c|c|c}
& \textrm{bc} & F_4 & SU(N_f = 3) & U(1)_a & U(1)_R \\ \hline
\textrm{VM} & \mathcal{N} & {\bf Adj} & {\bf 1} & 0 & 0 \\
Q_{\alpha} & \textrm{N} & {\bf 26} & {\bf 3} & 1 & 0 \\
 \hline
M_{\alpha \beta} & \textrm{N} & {\bf 1} & {\bf 6} & 2 & 0 \\
M_{\alpha \beta \gamma} & \textrm{N} & {\bf 1} & {\bf 10} & 3 & 0 \\
\overline{M}^{\alpha \beta} & \textrm{N} & {\bf 1} & {\bf \overline{6}} & 4 & 0 \\
\overline{M}^{\alpha} & \textrm{N} & {\bf 1} & {\bf \overline{3}} & 5 & 0 \\
\widetilde{M} & \textrm{N} & {\bf 1} & {\bf 1} & 6 & 0 \\
\widehat{M} & \textrm{N} & {\bf 1} & {\bf 1} & 9 & 0 \\
V & \textrm{D} & {\bf 1} & {\bf 1} & -18 & 2
\end{array}
\end{align}

The operators map with the chirals in theory A corresponding to gauge-invariant combinations of the fundamental chirals in theory A (e.g.\ $M_{\alpha \beta}$ and $M_{\alpha \beta \gamma}$ correspond to $Q_{\alpha}Q_{\beta}$ and $Q_{\alpha}Q_{\beta}Q_{\gamma}$ with the gauge indices contracted with the symmetric rank-$2$ and rank-$3$ $F_4$-invariant tensors) except for $V$ which is dual to the minimal monopole operator in theory A in the bulk.

With this choice of boundary conditions, we can easily check that the gauge anomaly cancels and that the 't Hooft anomalies match.
\begin{align}
\label{bdy_F4_3_anom_A}
\Acal^A = & \underbrace{9\Tr(s^2) + 26r^2}_{\textrm{VM}, \; \Ncal}
 - \underbrace{\left( 9\Tr(s^2) + 13\Tr(x^2) + 39(a-r)^2 \right)}_{Q_{\alpha}, \; N}
   \nonumber \\
  = & - 13\Tr(x^2) - 39a^2 + 78ar - 13r^2, \\
\label{bdy_F4_3_anom_B}
\Acal^B = & - \underbrace{\left( \frac{5}{2}\Tr(x^2) + 3(2a-r)^2 \right)}_{M_{\alpha \beta}, \; N}
   - \underbrace{\left( \frac{15}{2}\Tr(x^2) + 5(3a-r)^2 \right)}_{M_{\alpha \beta \gamma}, \; N}
   + \underbrace{\frac{1}{2}\big( -18a + r \big)^2}_{V, \; D}
  \nonumber \\
  & - \underbrace{\left( \frac{5}{2}\Tr(x^2) + 3(4a-r)^2 \right)}_{\overline{M}^{\alpha \beta}, \; N}
   - \underbrace{\left( \frac{1}{2}\Tr(x^2) + \frac{3}{2}(5a-r)^2 \right)}_{\overline{M}^{\alpha}, \; N}
   - \underbrace{\frac{1}{2}\big( 6a - r \big)^2}_{\widetilde{M}, \; N} - \underbrace{\frac{1}{2}\big( 9a - r \big)^2}_{\widehat{M}, \; N}
  \nonumber \\
  = & - 13 \Tr(x^2) - 39a^2 + 78ar - 13r^2. 
\end{align}

We then find the following half-index for theory A
\begin{align}
\label{bdy_F4_3_hindexA}
    \II^A_{\Ncal, N} = & \frac{(q)_{\infty}^4}{2^7 3^2} \prod_{i = 1}^4 \oint \frac{ds_i}{2\pi i s_i} \frac{\prod_{\mu \in \mathrm{Roots}} (s^{\mu}; q)_{\infty}}{\prod_{\alpha = 1}^3 (q^{r/2} a x_{\alpha}; q)_{\infty}^2 \prod_{\mu \in \mathrm{Short \; roots}} (q^{r/2} a x_{\alpha} s^{\mu}; q)_{\infty}}, 
\end{align}
where $\prod_{\alpha = 1}^3 x_{\alpha} = 1$ and
\begin{align}
    \prod_{\mu \in \mathrm{Roots}} (s^{\mu}; q)_{\infty} = & \left( \prod_{\mu \in \mathrm{Short \; roots}} (s^{\mu}; q)_{\infty} \right) \left( \prod_{\mu \in \mathrm{Long \; roots}} (s^{\mu}; q)_{\infty} \right), \\
    \prod_{\mu \in \mathrm{Short \; roots}} (s^{\mu}; q)_{\infty} = & \prod_{i < j}^4 (s^{\pm}_i s^{\pm}_j; q)_{\infty} (s^{\pm}_i s^{\mp}_j; q)_{\infty}, \\
    \prod_{\mu \in \mathrm{Long \; roots}} (s^{\mu}; q)_{\infty} = & \left( \prod_{i = 1}^4 (s^{\pm 2}_i; q)_{\infty} \right) \prod_{z_i \in \{-1, +1\}} (s_1^{z_1} s_2^{z_2} s_3^{z_3} s_4^{z_4}; q)_{\infty}. 
\end{align}
For $r=1/3$ and $x_{\alpha}=1$, we obtain
\begin{align}
\II^A_{\Ncal, N} = 1+6a^2q^{1/3}+10a^3q^{1/2}+27a^4q^{2/3}+63a^5q^{5/6}+\cdots. 
\end{align}

On the other hand, theory B has half-index
\begin{align}
\label{bdy_F4_3_hindexB}
    \II^B_{N^6; D} = & \frac{(q^{9r} a^{18};q)_{\infty}}{(q^{3r} a^6;q)_{\infty} (q^{9r/2} a^9;q)_{\infty} \left( \prod_{\alpha \le \beta}^3 (q^{r} a^2 x_{\alpha} x_{\beta}; q)_{\infty} \right) \prod_{\alpha = 1}^3 (q^{5r/2} a^5 x_{\alpha}^{-1};q)_{\infty}}
    \nonumber \\
    & \times \frac{1}{\left( \prod_{\alpha \le \beta}^3 (q^{2r} a^4 x_{\alpha}^{-1} x_{\beta}^{-1};q)_{\infty}\right) (q^{3r/2} a^3;q)_{\infty} \prod_{\alpha, \beta = 1}^3 (q^{3r/2} a^3 x_{\alpha}^2 x_{\beta};q)_{\infty}}. 
\end{align}
We have confirmed that this precisely coincides with the half-index (\ref{bdy_F4_3_hindexA}) of theory A. 

Ito \cite{MR2267266} proved an identity for $F_4$ which we can interpret as the matching of the above half-indices. 
In fact, we note that the contribution $\prod_{\alpha \le \beta}^3 (q^{2r} a^4 x_{\alpha}^{-1} x_{\beta}^{-1};q)_{\infty}$ from $\overline{M}^{\alpha \beta}$ in (\ref{bdy_F4_3_hindexB}) was erroneously replaced with a contribution corresponding to a fundamental representation of $SU(N_f = 3)$ in \cite{MR2267266}. 
This has been noted by Ito who has confirmed that the matching of (\ref{bdy_F4_3_hindexA}) and (\ref{bdy_F4_3_hindexB}) is the correct statement. 
\footnote{We thank Masahiko Ito and Masatoshi Noumi for useful communication regarding this issue. }

Switching all boundary conditions gives a Macdonald type sum which is independent of the gauge fugacities \cite{MR1868358, MR2016669}
\begin{align}
\label{bdy_F4_3_hindexA_D}
\II^A_{\Dcal, D} 
& =  \frac{1}{(q)_{\infty}^4} \sum_{m_i \in \Zb}  \frac{\prod_{\alpha = 1}^3 (q^{1 - r/2} a^{-1} x_{\alpha}^{-1}; q)_{\infty}^2 \prod_{\mu \in \mathrm{Short \; roots}} (q^{1 + m \cdot \mu - r/2} s^{\mu} a^{-1} x_{\alpha}^{-1}; q)_{\infty}} {\prod_{\mu \in \mathrm{Roots}} (q^{1 + m \cdot \mu} s^{\mu}; q)_{\infty}}
\end{align}
and matches the Dirchlet half-index of theory B
\begin{align}
\label{bdy_F4_3_hindexB_D}
\II^B_{D^6; N}
&= \frac{(q^{1 - 3r} a^{-6};q)_{\infty} (q^{1 - 9r/2} a^{-9};q)_{\infty} \left( \prod_{\alpha \le \beta}^3 (q^{1 - r} a^{-2} x_{\alpha}^{-1} x_{\beta}^{-1}; q)_{\infty} \right) \prod_{\alpha = 1}^3 (q^{1 - 5r/2} a^{-5} x_{\alpha};q)_{\infty}} {(q^{1 - 9r} a^{-18};q)_{\infty}}
    \nonumber \\
    & \times \left( \prod_{\alpha \le \beta}^3 (q^{1 - 2r} a^{-4} x_{\alpha} x_{\beta};q)_{\infty}\right)(q^{1 - 3r/2} a^{-3};q)_{\infty} \prod_{\alpha, \beta = 1}^3 (q^{1 - 3r/2} a^{-3} x_{\alpha}^{-2} x_{\beta}^{-1};q)_{\infty}. 
\end{align}
Again, this matching is equivalent to a known identity for a Macdonald type sum which is listed in \cite{MR2267266}.

%%%%%%%%%%%%%%%%%%%%%%%%%%%%%%%%%%
%%%%%%%%%%%%%%%%%%%%%%%%%%%%%%%%%%

%%%%%%%%%%%%%%%%%%%%%%%%%%%%%%%%%%
%%%%%%%%%%%%%%%%%%%%%%%%%%%%%%%%%%
\subsection{$F_4$ with an adjoint chiral}
\label{sec_F4_adj_integral}
%%%%%%%%%%%%%%%%%%%%%%%%%%%%%%%%%%
%%%%%%%%%%%%%%%%%%%%%%%%%%%%%%%%%%

The charges and boundary conditions are summarized as follows: 
\begin{align}
\label{F4_adj_charges}
\begin{array}{c|c|c|c|c}
& \textrm{bc} & F_4 & U(1)_a & U(1)_R \\ \hline
\textrm{VM} & \mathcal{N} & {\bf Adj} & 0 & 0 \\
\Phi & \textrm{N} & {\bf Adj} & 1 & 0 \\
 \hline
M_2 & \textrm{N} & {\bf 1} & 2 & 0 \\
M_6 & \textrm{N} & {\bf 1} & 6 & 0 \\
M_8 & \textrm{N} & {\bf 1} & 8 & 0 \\
M_{12} & \textrm{N} & {\bf 1} & 12 & 0 \\
V_1 & \textrm{D} & {\bf 1} & -1 & 0 \\
V_5 & \textrm{D} & {\bf 1} & -5 & 0 \\
V_7 & \textrm{D} & {\bf 1} & -7 & 0 \\
V_{11} & \textrm{D} & {\bf 1} & -11 & 0
\end{array}
\end{align}
The boundary 't Hooft anomaly of theory A is given by
\begin{align}
\label{bdy_F4_adj_anomA}
\Acal ^A& = \underbrace{{9\Tr(s^2)} + 26r^2}_{\textrm{VM}, \; \Ncal}
 - \underbrace{\left( 9\Tr(s^2) +26(a-r)^2 \right)}_{\Phi, \; N}
  \nonumber \\
  = & - 26a^2 + 52ar, 
\end{align}
which exactly agrees with the boundary anomaly of theory B which is given by
\begin{align}
\label{bdy_F4_adj_anomB}
\Acal ^B= & - \underbrace{\frac{1}{2}(2a-r)^2}_{M_2, \; N}
   + \underbrace{\frac{1}{2}( -a - r )^2}_{V_1, \; D}
   - \underbrace{\frac{1}{2}(6a-r)^2}_{M_6, \; N}
   + \underbrace{\frac{1}{2}( -5a - r )^2}_{V_5, \; D}
 \nonumber \\
    & - \underbrace{\frac{1}{2}(8a-r)^2}_{M_8, \; N}
   + \underbrace{\frac{1}{2}( -7a - r )^2}_{V_7, \; D}
   - \underbrace{\frac{1}{2}(12a-r)^2}_{M_{12}, \; N}
   + \underbrace{\frac{1}{2}( -11a - r )^2}_{V_{11}, \; D} . 
\end{align}

This gives the half-index for theory A
\begin{align}
\label{bdy_F4_adj_hindexA}
\mathbb{II}_{\mathcal{N},N}^A
&=\frac{(q)_{\infty}^4}{2^7 3^2} \prod_{i = 1}^4 \oint \frac{ds_i}{2\pi i s_i} 
\frac{\prod_{\mu \in \mathrm{Roots}} (s^{\mu}; q)_{\infty}}
{\prod_{\mu \in \mathrm{Roots}} (q^{r/2} a s^{\mu}; q)_{\infty}}. 
\end{align}

The Neumann half-index (\ref{bdy_F4_adj_hindexA}) agrees with the half-index for theory B
\begin{align}
\label{bdy_F4_adj_hindexB}
\mathbb{II}_{N^4,D^4}^B & =
\frac{
(q^{1+r/2}a;q)_{\infty} 
(q^{1+5r/2}a^5;q)_{\infty} 
(q^{1+7r/2}a^7;q)_{\infty} 
(q^{1+11r/2}a^{11};q)_{\infty} 
}
{
(q^{r}a^2;q)_{\infty} 
(q^{3r}a^3;q)_{\infty} 
(q^{4r}a^8;q)_{\infty} 
(q^{6r}a^{12};q)_{\infty} 
}, 
\end{align}
as shown in \cite{MR1314036}.

Switching all boundary conditions gives matching Dirchlet half-indices, equivalent to a known identity for a Macdonald type sum which is listed in \cite{MR2267266}
\begin{align}
\label{bdy_F4_adj_hindexA_D}
    \II^A_{\Dcal, D} = & \frac{1}{(q)_{\infty}^4} \sum_{m_i \in \Zb}  \frac{\prod_{\mu \in \mathrm{Roots}} (q^{1 + m \cdot \mu - r/2} s^{\mu} a^{-1}; q)_{\infty}} {\prod_{\mu \in \mathrm{Roots}} (q^{1 + m \cdot \mu} s^{\mu}; q)_{\infty}}
\end{align}
and
\begin{align}
\label{bdy_F4_adj_hindexB_D}
\II^B_{D^4, N^4} = & 
\frac
{
(q^{1-r}a^{-2};q)_{\infty} 
(q^{1-3r}a^{-3};q)_{\infty} 
(q^{1-4r}a^{-8};q)_{\infty} 
(q^{1-6r}a^{-12};q)_{\infty} 
}
{
(q^{-r/2}a^{-1};q)_{\infty} 
(q^{-5r/2}a^{-5};q)_{\infty} 
(q^{-7r/2}a^{-7};q)_{\infty} 
(q^{-11r/2}a^{-11};q)_{\infty} 
}. 
\end{align}

%%%%%%%%%%%%%%%%%%%%%%%%%%%%%%%%%%
%%%%%%%%%%%%%%%%%%%%%%%%%%%%%%%%%%
\section{Gauge group $E_6$}
\label{sec_E6}
%%%%%%%%%%%%%%%%%%%%%%%%%%%%%%%%%%
%%%%%%%%%%%%%%%%%%%%%%%%%%%%%%%%%%
In this section we consider boundary confining dualities for $E_6$ gauge theories with (anti)fundamental or adjoint chirals. The results here for the cases with fundamentals are new boundary duality results for known bulk dualities discussed in \cite{Nii:2019dwi}. The conjectured matching of half-indices gives new examples of generalized Askey-Wilson and Macdonald sum identities for $E_6$ which have not been studied in the mathematical literature. We give some numerical evidence for these conjectured boundary confining dualities.

%%%%%%%%%%%%%%%%%%%%%%%%%%%%%%%%%%
%%%%%%%%%%%%%%%%%%%%%%%%%%%%%%%%%%
\subsection{$E_6$ with $4$ fundamental chirals}
\label{sec_E6_4_integral}
%%%%%%%%%%%%%%%%%%%%%%%%%%%%%%%%%%
%%%%%%%%%%%%%%%%%%%%%%%%%%%%%%%%%%

We have dual confining boundary conditions for $E_6$ gauge theory with $4$ fundamental chirals
\begin{align}
\label{E6_4_charges}
\begin{array}{c|c|c|c|c|c}
& \textrm{bc} & E_6 & SU(N_f = 4) & U(1)_a & U(1)_R \\ \hline
\textrm{VM} & \mathcal{N} & {\bf Adj} & {\bf 1} & 0 & 0 \\
Q_{\alpha} & \textrm{N} & {\bf 27} & {\bf 4} & 1 & 0 \\
 \hline
M_{\alpha \beta \gamma} & \textrm{N} & {\bf 1} & {\bf 20} & 3 & 0 \\
\overline{M}^{\alpha \beta} & \textrm{N} & {\bf 1} & {\bf \overline{10}} & 6 & 0 \\
M & \textrm{N} & {\bf 1} & {\bf 1} & 12 & 0 \\
V & \textrm{D} & {\bf 1} & {\bf 1} & -24 & 2
\end{array}
\end{align}

We see that the anomaly polynomials match for theories A and B:
\begin{align}
\label{bdy_E6_4_anom_A}
\Acal^A = & \underbrace{12\Tr(s^2) + 39r^2}_{\textrm{VM}, \; \Ncal}
 - \underbrace{\left( 12\Tr(s^2) + \frac{27}{2} \Tr(x^2) + 54(a-r)^2 \right)}_{Q_{\alpha}, \; N}
   \nonumber \\
  = & - \frac{27}{2} \Tr(x^2) - 54a^2 + 108ar - 15r^2, \\
\label{bdy_E6_4_anom_B}
\Acal^B = & - \underbrace{\left( \frac{21}{2}\Tr(x^2) + 10(3a-r)^2 \right)}_{M_{\alpha \beta \gamma}, \; N}
   + \underbrace{\frac{1}{2}\big( -24a + r \big)^2}_{V, \; D}
  \nonumber \\
  & - \underbrace{\left( 3\Tr(x^2) + 5(6a-r)^2 \right)}_{\overline{M}^{\alpha \beta}, \; N}
 - \underbrace{\frac{1}{2}\big( 12a - r \big)^2}_{M, \; N}
  \nonumber \\
  = & - \frac{27}{2} \Tr(x^2) - 54a^2 + 108ar - 15r^2. 
\end{align}

Similarly the half-indices match. Theory A has half-index
\begin{align}
    \II^A_{\Ncal, N} = & \frac{(q)_{\infty}^6}{2^7 3^4 5} \prod_{n = 1}^3 \prod_{i = 1}^2 \oint \frac{ds_{n,i}}{2\pi i s_{n,i}} \frac{\prod_{\mu \in \mathrm{Roots}} (s^{\mu}; q)_{\infty}}{\prod_{\alpha = 1}^4 \prod_{\mu \in \mathrm{Fund. \; weights}} (q^{r/2} s^{\mu} a x_{\alpha}; q)_{\infty}}, 
\end{align}
where $\prod_{\alpha = 1}^4 x_{\alpha} = 1$ and
\begin{align}
    \prod_{\mu \in \mathrm{Roots}} (s^{\mu}; q)_{\infty} = & \left( \prod_{i, j, k = 1}^3 (s_{1, i}^{\pm} s_{2, j}^{\pm} s_{3, k}^{\pm}; q)_{\infty} \right) \prod_{n = 1}^3 \prod_{i < j}^3 (s^{\pm}_{n,i} s^{\mp}_{n,j}; q)_{\infty}, \\
    \prod_{\mu \in \mathrm{Fund. \; weights}} (s^{\mu}; q)_{\infty} = & \prod_{i, j = 1}^3 (s_{1, i} s_{2, j}^{-1}; q)_{\infty} \prod_{i, j = 1}^3 (s_{2, i} s_{3, j}^{-1}; q)_{\infty} \prod_{i, j = 1}^3 (s_{3, i} s_{1, j}^{-1}; q)_{\infty}, 
\end{align}
with $\prod_{i = 1}^3 s_{n, i} = 1$. Theory B has half-index
\begin{align}
    \II^B_{N, N, N, D} = & \frac{(q^{12r} a^{24};q)_{\infty}}{(q^{6r} a^{12};q)_{\infty} \left( \prod_{\alpha \le \beta}^4 (q^{3r} a^6 x_{\alpha}^{-1} x_{\beta}^{-1}; q)_{\infty} \right) \prod_{\alpha \le \beta \le \gamma}^4 (q^{3r/2} a^3 x_{\alpha} x_{\beta} x_{\gamma};q)_{\infty}}. 
\end{align}

Calculating the $q$-series expansion of the theory A half-index is computationally expensive. However, we can consider the reduced half-index given by taking the $q \to 0$ limit after fixing $\mathfrak{t} = q^{r/2} a$ and $x_{\alpha}$. It is then straightforward to numerically check the evaluation of the theory A reduced half-index for various choices of $\mathfrak{t}$ and $x_{\alpha}$ such that $|\mathfrak{t} x_{\alpha}| < 1$. We have checked that the Mathematica Monte Carlo numerical integration matches the evaluation of the theory B reduced half-index to the order of approximately $1\%$, providing strong numerical evidence for the claimed boundary confining duality. Details are given in appendix~\ref{MC_int}.

Switching all boundary conditions gives
\begin{align}
    \II^A_{\Dcal, D} = & \frac{1}{(q)_{\infty}^6} \sum_{m_{n, i} \in \Zb} \frac{\prod_{\alpha = 1}^4 \prod_{\mu \in \mathrm{Fund. \; weights}} (q^{1 - m \cdot \mu - r/2} s^{-\mu} a^{-1} x_{\alpha}; q)_{\infty}} {\prod_{\mu \in \mathrm{Roots}} (q^{1 + m \cdot \mu} s^{\mu}; q)_{\infty}},
\end{align}
with $\sum_{i = 1}^3 m_{n, i} = 0$. Theory B has half-index
\begin{align}
    \II^B_{D, D, D, N} = & \frac{(q^{1 - 6r} a^{-12};q)_{\infty} \left( \prod_{\alpha \le \beta}^4 (q^{1 - 3r} a^{-6} x_{\alpha} x_{\beta}; q)_{\infty} \right) \prod_{\alpha \le \beta \le \gamma}^4 (q^{1 - 3r/2} a^{-3} x_{\alpha}^{-1} x_{\beta}^{-1} x_{\gamma}^{-1};q)_{\infty}} {(q^{1 - 12r} a^{-24};q)_{\infty}}. 
\end{align}
We conjecture a new Macdonald type sum identity corresponding to the matching of these half-indices.

%%%%%%%%%%%%%%%%%%%%%%%%%%%%%%%%%%
%%%%%%%%%%%%%%%%%%%%%%%%%%%%%%%%%%
\subsection{$E_6$ with $3$ fundamental and $1$ antifundamental chirals}
\label{sec_E6_3p1_integral}
%%%%%%%%%%%%%%%%%%%%%%%%%%%%%%%%%%
%%%%%%%%%%%%%%%%%%%%%%%%%%%%%%%%%%

We have dual confining boundary conditions for $E_6$ gauge theory with $3$ fundamental chirals and $1$ antifundamental chiral
\begin{align}
\label{E6_3p1_charges}
\begin{array}{c|c|c|c|c|c|c}
& \textrm{bc} & E_6 & SU(N_f = 3) & U(1)_a & U(1)_b & U(1)_R \\ \hline
\textrm{VM} & \mathcal{N} & {\bf Adj} & {\bf 1} & 0 & 0 & 0 \\
Q_{\alpha} & \textrm{N} & {\bf 27} & {\bf 3} & 1 & 0 & 0 \\
\overline{Q} & \textrm{N} & {\bf \overline{27}} & {\bf 1} & 0 & 1 & 0 \\
 \hline
M_{\alpha} & \textrm{N} & {\bf 1} & {\bf 3} & 1 & 1 & 0 \\
B_{\alpha \beta \gamma} & \textrm{N} & {\bf 1} & {\bf 10} & 3 & 0 & 0 \\
\overline{B} & \textrm{N} & {\bf 1} & {\bf 1} & 0 & 3 & 0 \\
M_{\alpha \beta} & \textrm{N} & {\bf 1} & {\bf 6} & 2 & 2 & 0 \\
\overline{M}^{\alpha \beta} & \textrm{N} & {\bf 1} & {\bf \overline{6}} & 4 & 1 & 0 \\
M & \textrm{N} & {\bf 1} & {\bf 1} & 6 & 0 & 0 \\
\widetilde{M}^{\alpha \beta} & \textrm{N} & {\bf 1} & {\bf \overline{3}} & 5 & 2 & 0 \\
\widetilde{M} & \textrm{N} & {\bf 1} & {\bf 1} & 9 & 3 & 0 \\
V & \textrm{D} & {\bf 1} & {\bf 1} & -18 & -6 & 2
\end{array}
\end{align}

We see that the anomaly polynomials match for theories A and B:
\begin{align}
\label{bdy_E6_3p1_anom_A}
\Acal^A = & \underbrace{12\Tr(s^2) + 39r^2}_{\textrm{VM}, \; \Ncal}
 - \underbrace{\left( 9\Tr(s^2) + \frac{27}{2} \Tr(x^2) +  \frac{81}{2} (a-r)^2 \right)}_{Q_{\alpha}, \; N}
   \nonumber \\
 & - \underbrace{\left( 3\Tr(s^2) +  \frac{27}{2} (b-r)^2 \right)}_{Q_{\alpha}, \; N}
   \nonumber \\
  = & - \frac{27}{2} \Tr(x^2) - \frac{81}{2} a^2 + 81ar - \frac{27}{2} b^2 + 27br - 15r^2, \\
\label{bdy_E6_3p1_anom_B}
\Acal^B = & - \underbrace{\left( \frac{1}{2}\Tr(x^2) + \frac{3}{2}(a + b - r)^2 \right)}_{M_{\alpha}, \; N}
 - \underbrace{\left( \frac{15}{2}\Tr(x^2) + 10(3a-r)^2 \right)}_{B_{\alpha \beta \gamma}, \; N}
 - \underbrace{\frac{1}{2} (3b - r)^2}_{\overline{B}, \; N}
  \nonumber \\
  & - \underbrace{\left( \frac{5}{2}\Tr(x^2) + 3(2a + 2b - r)^2 \right)}_{M_{\alpha \beta}, \; N}
 - \underbrace{\left( \frac{5}{2}\Tr(x^2) + 3(4a + b - r)^2 \right)}_{\overline{M}^{\alpha \beta}, \; N}
 - \underbrace{\frac{1}{2}\big( 6a - r \big)^2}_{M, \; N}
   \nonumber \\
  & - \underbrace{\left( \frac{1}{2}\Tr(x^2) + \frac{3}{2}(5a + 2b - r)^2 \right)}_{\widetilde{M}^{\alpha \beta}, \; N}
 - \underbrace{\frac{1}{2} (9a + 3b - r)^2}_{\widetilde{M}, \; N}
  + \underbrace{\frac{1}{2} (-18a -6b + r)^2}_{V, \; D}
  \nonumber \\
  = & - \frac{27}{2} \Tr(x^2) - \frac{81}{2} a^2 + 81ar - \frac{27}{2} b^2 + 27br - 15r^2. 
\end{align}

Similarly the half-indices match. Theory A has half-index
\begin{align}
    \II^A_{\Ncal, N, N} = & \frac{(q)_{\infty}^6}{2^7 3^4 5} \prod_{n = 1}^3 \prod_{i = 1}^2 \oint \frac{ds_{n,i}}{2\pi i s_{n,i}} \frac{\prod_{\mu \in \mathrm{Roots}} (s^{\mu}; q)_{\infty}}{\prod_{\mu \in \mathrm{Fund. \; weights}} (q^{r_b/2} s^{-\mu} b; q)_{\infty} \prod_{\alpha = 1}^3 (q^{r_a/2} s^{\mu} a x_{\alpha}; q)_{\infty}}, 
\end{align}
where $\prod_{\alpha = 1}^3 x_{\alpha} = 1$ and
\begin{align}
    \prod_{\mu \in \mathrm{Roots}} (s^{\mu}; q)_{\infty} = & \left( \prod_{i, j, k = 1}^3 (s_{1, i}^{\pm} s_{2, j}^{\pm} s_{3, k}^{\pm}; q)_{\infty} \right) \prod_{n = 1}^3 \prod_{i < j}^3 (s^{\pm}_{n,i} s^{\mp}_{n,j}; q)_{\infty}, \\
    \prod_{\mu \in \mathrm{Fund. \; weights}} (s^{\mu}; q)_{\infty} = & \prod_{i, j = 1}^3 (s_{1, i} s_{2, j}^{-1}; q)_{\infty} \prod_{i, j = 1}^3 (s_{2, i} s_{3, j}^{-1}; q)_{\infty} \prod_{i, j = 1}^3 (s_{3, i} s_{1, j}^{-1}; q)_{\infty}, 
\end{align}
with $\prod_{i = 1}^3 s_{n, i} = 1$. Theory B has half-index
\begin{align}
    \II^B_{N^8; D} = & \frac{(q^{9r_a + 3r_b} a^{18}b^6;q)_{\infty}}{(q^{3r_b/2} b^3;q)_{\infty} (q^{3r_a} a^6;q)_{\infty} (q^{9r_a/2 + 3r_b/2} a^9 b^3;q)_{\infty}}
    \nonumber \\
    & \times \frac{1}{\prod_{\alpha \le \beta}^3 (q^{r_a + r_b} a^2 b^2 x_{\alpha} x_{\beta}; q)_{\infty} (q^{2r_a + r_b/2} a^4 b x_{\alpha}^{-1} x_{\beta}^{-1}; q)_{\infty}}
    \nonumber \\
    & \times \frac{1}{\left( \prod_{\alpha \le \beta \le \gamma}^3 (q^{3r_a/2} a^3 x_{\alpha} x_{\beta} x_{\gamma};q)_{\infty} \right) \prod_{\alpha = 1}^3 (q^{r_a/2 + r_b/2} a b x_{\alpha}; q)_{\infty} (q^{5r_a/2 + r_b} a^5 b^2 x_{\alpha}^{-1}; q)_{\infty}}. 
\end{align}

Again, calculating the $q$-series expansion of the theory A half-index is computationally expensive. However, we can consider the reduced half-index given by taking the $q \to 0$ limit after fixing $\mathfrak{t} = q^{r/2} a$ and $x_{\alpha}$. We have checked that the Mathematica Monte Carlo numerical integration of the theory A reduced half-index matches the evaluation of the theory B reduced half-index for various choices of $\mathfrak{t}$ and $x_{\alpha}$ such that $|\mathfrak{t} x_{\alpha}| < 1$ to the order of approximately $1\%$ as shown in appendix~\ref{MC_int}, providing strong numerical evidence for the claimed boundary confining duality.

Switching all boundary conditions gives
\begin{align}
    \II^A_{\Dcal, D, D} = & \frac{1}{(q)_{\infty}^6} \sum_{m_{n, i} \in \Zb} \frac{\prod_{\mu \in \mathrm{Fund. \; weights}} (q^{1 + m \cdot \mu - r_b/2} s^{\mu} b^{-1}; q)_{\infty} \prod_{\alpha = 1}^3 (q^{1 - m \cdot \mu - r_a/2}s^{-\mu} a^{-1} x_{\alpha}^{-1}; q)_{\infty}} {\prod_{\mu \in \mathrm{Roots}} (q^{1 + m \cdot \mu} s^{\mu}; q)_{\infty}}, 
\end{align}
with $\sum_{i = 1}^3 m_{n, i} = 0$ and theory B has half-index
\begin{align}
    \II^B_{D^8; N} = & \frac{(q^{1 - 3r_b/2} b^{-3};q)_{\infty} (q^{1 - 3r_a} a^{-6};q)_{\infty} (q^{1 - 9r_a/2 - 3r_b/2} a^{-9} b^{-3};q)_{\infty}} {(q^{1 - 9r_a - 3r_b} a^{-18}b^{-6};q)_{\infty}}
    \nonumber \\
    & \times \prod_{\alpha \le \beta}^3 (q^{1 - r_a - r_b} a^{-2} b^{-2} x_{\alpha}^{-1} x_{\beta}^{-1}; q)_{\infty} (q^{1 - 2r_a - r_b/2} a^{-4} b^{-1} x_{\alpha} x_{\beta}; q)_{\infty}
    \nonumber \\
    & \times \left( \prod_{\alpha \le \beta \le \gamma}^3 (q^{1 - 3r_a/2}a^3 x_{\alpha}^{-1} x_{\beta}^{-1} x_{\gamma}^{-1};q)_{\infty} \right) 
    \nonumber \\
    &\times \prod_{\alpha = 1}^3 (q^{1 - r_a/2 - r_b/2} a^{-1} b^{-1} x_{\alpha}^{-1}; q)_{\infty} (q^{1 - 5r_a/2 - r_b} a^{-5} b^{-2} x_{\alpha}; q)_{\infty}. 
\end{align}
We conjecture a new Macdonald type sum identity corresponding to the matching of these half-indices.

%%%%%%%%%%%%%%%%%%%%%%%%%%%%%%%%%%
%%%%%%%%%%%%%%%%%%%%%%%%%%%%%%%%%%
\subsection{$E_6$ with $2$ fundamental and $2$ antifundamental chirals}
\label{sec_E6_2p2_integral}
%%%%%%%%%%%%%%%%%%%%%%%%%%%%%%%%%%
%%%%%%%%%%%%%%%%%%%%%%%%%%%%%%%%%%

We have dual confining boundary conditions for $E_6$ gauge theory with $2$ fundamental chirals and $2$ antifundamental chirals
\begin{align}
\label{E6_2p2_charges}
\begin{array}{c|c|c|c|c|c|c|c}
& \textrm{bc} & E_6 & SU(N_f = 2) & SU(N_a = 2) & U(1)_a & U(1)_b & U(1)_R \\ \hline
\textrm{VM} & \mathcal{N} & {\bf Adj} & {\bf 1} & {\bf 1} & 0 & 0 & 0 \\
Q_I & \textrm{N} & {\bf 27} & {\bf 2} & {\bf 1} & 1 & 0 & 0 \\
\overline{Q}_{\alpha} & \textrm{N} & {\bf \overline{27}} & {\bf 1} & {\bf 2} & 0 & 1 & 0 \\
 \hline
M_{I \alpha} & \textrm{N} & {\bf 1} & {\bf 2} & {\bf 2} & 1 & 1 & 0 \\
B_{IJK} & \textrm{N} & {\bf 1} & {\bf 4} & {\bf 1} & 3 & 0 & 0 \\
\widetilde{B}_{\alpha \beta \gamma} & \textrm{N} & {\bf 1} & {\bf 1} & {\bf 4} & 0 & 3 & 0 \\
M_{IJ \alpha \beta} & \textrm{N} & {\bf 1} & {\bf 3} & {\bf 3} & 2 & 2 & 0 \\
M_I & \textrm{N} & {\bf 1} & {\bf 2} & {\bf 1} & 1 & 4 & 0 \\
M_{\alpha} & \textrm{N} & {\bf 1} & {\bf 1} & {\bf 2} & 4 & 1 & 0 \\
\widehat{M}_{I \alpha} & \textrm{N} & {\bf 1} & {\bf 2} & {\bf 2} & 3 & 3 & 0 \\
M & \textrm{N} & {\bf 1} & {\bf 1} & {\bf 1} & 4 & 4 & 0 \\
\widehat{M} & \textrm{N} & {\bf 1} & {\bf 1} & {\bf 1} & 6 & 6 & 0 \\
V & \textrm{D} & {\bf 1} & {\bf 1} & {\bf 1} & -12 & -12 & 2
\end{array}
\end{align}

We see that the anomaly polynomials match for theories A and B:
\begin{align}
\label{bdy_E6_2p2_anom_A}
\Acal^A = & \underbrace{12\Tr(s^2) + 39r^2}_{\textrm{VM}, \; \Ncal}
 - \underbrace{\left( 6\Tr(s^2) + \frac{27}{2} \Tr(x^2) +  \frac{54}{2} (a-r)^2 \right)}_{Q_I, \; N}
   \nonumber \\
 & - \underbrace{\left( 6\Tr(s^2) + \frac{27}{2} \Tr(\tilde{x}^2) +  \frac{54}{2} (b-r)^2 \right)}_{\widetilde{Q}_{\alpha}, \; N}
   \nonumber \\
  = & - \frac{27}{2} \Tr(x^2) - \frac{27}{2} \Tr(\tilde{x}^2) - 27a^2 + 54ar - 27b^2 + 54br - 15r^2, \\
\label{bdy_E6_2p2_anom_B}
\Acal^B = & - \underbrace{\left( \Tr(x^2) + \Tr(\tilde{x}^2) + 2(a + b - r)^2 \right)}_{M_{I \alpha}, \; N}
 - \underbrace{\left( 5\Tr(x^2) + 2(3a-r)^2 \right)}_{B_{IJK}, \; N}
  - \underbrace{\left( 5\Tr(\tilde{x}^2) + 2(3b-r)^2 \right)}_{\widetilde{B}_{\alpha \beta \gamma}, \; N}
  \nonumber \\
  & - \underbrace{\left( 6\Tr(x^2) + 6\Tr(\tilde{x}^2) + \frac{9}{2}(2a + 2b - r)^2 \right)}_{M_{IJ \alpha \beta}, \; N}
 - \underbrace{\left( \frac{1}{2}\Tr(x^2) + (a + 4b - r)^2 \right)}_{M_I, \; N}
    \nonumber \\
  & - \underbrace{\left( \frac{1}{2}\Tr(\tilde{x}^2) + (4a + b - r)^2 \right)}_{M_{\alpha}, \; N}
  - \underbrace{\left( \Tr(x^2) + \Tr(\tilde{x}^2) + 2(3a + 3b - r)^2 \right)}_{\widehat{M}_{I \alpha}, \; N}
   \nonumber \\
  & - \underbrace{\frac{1}{2} (4a + 4b - r)^2}_{M, \; N}
  - \underbrace{\frac{1}{2} (6a + 6b - r)^2}_{\widehat{M}, \; N}
  + \underbrace{\frac{1}{2} (-12a - 12b + r)^2}_{V, \; D}
  \nonumber \\
  = & - \frac{27}{2} \Tr(x^2) - \frac{27}{2} \Tr(\tilde{x}^2) - 27a^2 + 54ar - 27b^2 + 54br - 15r^2. 
\end{align}

Similarly the half-indices match. Theory A has half-index
\begin{align}
    \II^A_{\Ncal, N, N} = & \frac{(q)_{\infty}^6}{2^7 3^4 5} \prod_{n = 1}^3 \prod_{i = 1}^2 \oint \frac{ds_{n,i}}{2\pi i s_{n,i}} \frac{\prod_{\mu \in \mathrm{Roots}} (s^{\mu}; q)_{\infty}}{\prod_{\mu \in \mathrm{Fund. \; weights}} \left( \prod_{I = 1}^2 (q^{r_a/2} s^{\mu} a x_I; q)_{\infty} \right) \left( \prod_{\alpha = 1}^2 (q^{r_b/2} s^{-\mu} b \tilde{x}_{\alpha}; q)_{\infty} \right)}, 
\end{align}
where $\prod_{I = 1}^2 x_I = \prod_{\alpha = 1}^2 \tilde{x}_{\alpha} = 1$ and
\begin{align}
    \prod_{\mu \in \mathrm{Roots}} (s^{\mu}; q)_{\infty} = & \left( \prod_{i, j, k = 1}^3 (s_{1, i}^{\pm} s_{2, j}^{\pm} s_{3, k}^{\pm}; q)_{\infty} \right) \prod_{n = 1}^3 \prod_{i < j}^3 (s^{\pm}_{n,i} s^{\mp}_{n,j}; q)_{\infty}, \\
    \prod_{\mu \in \mathrm{Fund. \; weights}} (s^{\mu}; q)_{\infty} = & \prod_{i, j = 1}^3 (s_{1, i} s_{2, j}^{-1}; q)_{\infty} \prod_{i, j = 1}^3 (s_{2, i} s_{3, j}^{-1}; q)_{\infty} \prod_{i, j = 1}^3 (s_{3, i} s_{1, j}^{-1}; q)_{\infty}, 
\end{align}
with $\prod_{i = 1}^3 s_{n, i} = 1$. Theory B has half-index
\begin{align}
    \II^B_{N^9; D} = & \frac{(q^{6r_a + 6r_b} a^{12}b^{12};q)_{\infty}}{(q^{2r_a + 2r_b} a^4 b^4;q)_{\infty} (q^{3r_a + 3r_b} a^6 b^6;q)_{\infty} \left( \prod_{I \le J}^2 \prod_{\alpha \le \beta}^2 (q^{r_a + r_b} a^2 b^2 x_I x_J \tilde{x}_{\alpha} \tilde{x}_{\beta}; q)_{\infty} \right) }
    \nonumber \\
    & \times \frac{1}{\prod_{I = 1}^2 (q^{r_a/2 + 2r_b} a b^4 x_I; q)_{\infty} \prod_{\alpha = 1}^2 (q^{2r_a + r_b/2} a^4 b \tilde{x}_{\alpha}; q)_{\infty} }
    \nonumber \\
    & \times \frac{1}{\left( \prod_{I \le J \le K}^2 (q^{3r_a/2} a^3 x_I x_J x_K;q)_{\infty} \right) \left( \prod_{\alpha \le \beta \le \gamma}^2 (q^{3r_b}/2 b^3 \tilde{x}_{\alpha} \tilde{x}_{\beta} \tilde{x}_{\gamma};q)_{\infty} \right)}
    \nonumber \\
    & \times \frac{1}{\prod_{I = 1}^2 \prod_{\alpha = 1}^2 (q^{r_a/2 + r_b/2} a b x_I \tilde{x}_{\alpha}; q)_{\infty} (q^{3r_a/2 + 3r_b/2}a^3 b^3 x_I \tilde{x}_{\alpha}; q)_{\infty}}. 
\end{align}

As for the previous cases with $E_6$ gauge group, calculating the $q$-series expansion of the theory A half-index is computationally expensive. However, we can consider the reduced half-index given by taking the $q \to 0$ limit after fixing $\mathfrak{t} = q^{r/2} a$ and $x_{\alpha}$. We have checked that the Mathematica Monte Carlo numerical integration of the theory A reduced half-index matches the evaluation of the theory B reduced half-index for various choices of $\mathfrak{t}$ and $x_{\alpha}$ such that $|\mathfrak{t} x_{\alpha}| < 1$ to the order of approximately $1\%$ as shown in appendix~\ref{MC_int}, providing strong numerical evidence for the claimed boundary confining duality.

Switching all boundary conditions gives
\begin{align}
    \II^A_{\Dcal, D, D} = & \frac{1}{(q)_{\infty}^6} \sum_{m_{n, i} \in \Zb} \frac{\prod_{\mu \in \mathrm{Fund. \; weights}} \prod_{I = 1}^2 (q^{1 - m \cdot \mu - r_a/2} s^{-\mu} a^{-1} x_I^{-1}; q)_{\infty}} {\prod_{\mu \in \mathrm{Roots}} (q^{1 + m \cdot \mu} s^{\mu}; q)_{\infty}} 
 \nonumber \\
    & \times \prod_{\mu \in \mathrm{Fund. \; weights}} \prod_{\alpha = 1}^2 (q^{1 + m \cdot \mu - r_b/2}s^{\mu} b^{-1} \tilde{x}_{\alpha}^{-1}; q)_{\infty}, 
\end{align}
where $\sum_{i = 1}^3 m_{n, i} = 0$. Theory B has half-index
\begin{align}
    \II^B_{D^9; N} = & \frac{(q^{1 - 2r_A - 2r_b} a^{-4} b^{-4};q)_{\infty} (q^{1 - 3r_A - 3r_b} a^{-6} b^{-6};q)_{\infty} \prod_{I = 1}^2 (q^{1 - r_A/2 - 2r_b} a^{-1} b^{-4} x_I^{-1}; q)_{\infty}} {(q^{1 - 6r_A - 6r_b} a^{-12}b^{-12};q)_{\infty}}
    \nonumber \\
    & \times \left( \prod_{I \le J}^2 \prod_{\alpha \le \beta}^2 (q^{1 - r_A - r_b} a^{-2} b^{-2} x_I^{-1} x_J^{-1} \tilde{x}_{\alpha}^{-1} \tilde{x}_{\beta}^{-1}; q)_{\infty} \right)
    \prod_{\alpha = 1}^2 (q^{1 - 2r_A - r_b/2} a^{-4} b^{-1} \tilde{x}_{\alpha}^{-1}; q)_{\infty}
    \nonumber \\
    & \times \left( \prod_{I \le J \le K}^2 (q^{1 - 3r_A/2} a^{-3} x_I^{-1} x_J^{-1} x_K^{-1};q)_{\infty} \right) \prod_{\alpha \le \beta \le \gamma}^2 (q^{1 - 3r_b/2} b^{-3} \tilde{x}_{\alpha}^{-1} \tilde{x}_{\beta}^{-1} \tilde{x}_{\gamma}^{-1};q)_{\infty}
    \nonumber \\
    & \times \prod_{I = 1}^2 \prod_{\alpha = 1}^2 (q^{1 - r_A/2 - r_b/2} a^{-1} b^{-1} x_I^{-1} \tilde{x}_{\alpha}^{-1}; q)_{\infty} (q^{1 - 3r_A/2 - 3r_b/2} a^{-3} b^{-3} x_I^{-1} \tilde{x}_{\alpha}^{-1}; q)_{\infty}. 
\end{align}
We conjecture a new Macdonald type sum identity corresponding to the matching of these half-indices.

%%%%%%%%%%%%%%%%%%%%%%%%%%%%%%%%%%
%%%%%%%%%%%%%%%%%%%%%%%%%%%%%%%%%%

%%%%%%%%%%%%%%%%%%%%%%%%%%%%%%%%%%
%%%%%%%%%%%%%%%%%%%%%%%%%%%%%%%%%%
\subsection{$E_6$ with an adjoint chiral}
\label{sec_E6_adj_integral}
%%%%%%%%%%%%%%%%%%%%%%%%%%%%%%%%%%
%%%%%%%%%%%%%%%%%%%%%%%%%%%%%%%%%%

The charges and boundary conditions are summarized as follows: 
\begin{align}
\label{E6_adj_charges}
\begin{array}{c|c|c|c|c}
& \textrm{bc} & E_6 & U(1)_a & U(1)_R \\ \hline
\textrm{VM} & \mathcal{N} & {\bf Adj} & 0 & 0 \\
\Phi & \textrm{N} & {\bf Adj} & 1 & 0 \\
 \hline
M_I & \textrm{N} & {\bf 1} & I & 0 \\
V_{I-1} & \textrm{D} & {\bf 1} & -(I-1) & 0
\end{array}
\end{align}
where $I \in \{2, 5, 6, 8, 9, 12\}$.

The boundary 't Hooft anomaly of theory A is given by
\begin{align}
\label{bdy_E6_adj_anomA}
\Acal ^A& = \underbrace{{12\Tr(s^2)} + 39r^2}_{\textrm{VM}, \; \Ncal}
 - \underbrace{\left( 12\Tr(s^2) + 39(a-r)^2 \right)}_{\Phi, \; N}
  \nonumber \\
  = & - 39a^2 + 78ar , 
\end{align}
which exactly agrees with the boundary anomaly of theory B which is given by
\begin{align}
\label{bdy_E6_adj_anomB}
\Acal ^B= & \sum_{I \in \{2, 5, 6, 8, 9, 12\}} \left(
   - \underbrace{\frac{1}{2}\big(Ia - r\big)^2}_{M_I, \; N}
   + \underbrace{\frac{1}{2}\big((1-I)a - r\big)^2}_{V_{I-1}, \; D} \right) . 
\end{align}

This gives the half-index for theory A
\begin{align}
\label{bdy_E6_adj_hindexA}
    \II^A_{\Ncal, N} = & \frac{1}{2^7 \cdot 3^4 \cdot 5}
    \frac{(q)_{\infty}^6}{(q^{r/2} a; q)_{\infty}^6}
    \prod_{n = 1}^3 \prod_{i = 1}^2 \oint \frac{ds_{n,i}}{2\pi i s_{n,i}} \frac{\prod_{\mu \in \mathrm{Roots}} (s^{\mu}; q)_{\infty}}{\prod_{\mu \in \mathrm{Roots}} (q^{r/2} a s^{\mu}; q)_{\infty}}, 
\end{align}
where
\begin{align}
    \prod_{\mu \in \mathrm{Roots}} (s^{\mu}; q)_{\infty} = & \left( \prod_{i, j, k = 1}^3 (s_{1, i}^{\pm} s_{2, j}^{\pm} s_{3, k}^{\pm}; q)_{\infty} \right) \prod_{n = 1}^3 \prod_{i < j}^3 (s^{\pm}_{n,i} s^{\mp}_{n,j}; q)_{\infty}, 
\end{align}
with $\prod_{i = 1}^3 s_{n,i} = 1$.

The half-index (\ref{bdy_E6_adj_hindexA}) coincides \cite{MR1314036} with the half-index for theory B
\begin{align}
\label{bdy_E6_adj_hindexB}
\mathbb{II}_{N^6; D^6}^{B}
&=
\frac{
(q^{1+r/2}a;q)_{\infty}
(q^{1+4r/2}a^4;q)_{\infty}
(q^{1+5r/2}a^5;q)_{\infty}
}
{
(q^{r}a^{2};q)_{\infty}
(q^{5r/2}a^{5};q)_{\infty}
(q^{3r} a^{6};q)_{\infty}
}
\nonumber\\
&\times 
\frac{
(q^{1+7r/2}a^{7};q)_{\infty}
(q^{1+4r}a^{8};q)_{\infty}
(q^{1+11r/2}a^{11};q)_{\infty}
}
{
(q^{4r}a^{8};q)_{\infty}
(q^{9r/2}a^{9};q)_{\infty}
(q^{6r}a^{12};q)_{\infty}
}.
\end{align}

Switching all boundary conditions gives
\begin{align}
\label{bdy_E6_adj_hindexA_D}
    \II^A_{\Dcal, D} = & 
    \frac{(q^{1 - r/2} a^{-1}; q)_{\infty}^6}{(q)_{\infty}^6}
    \sum_{m_{n, i} \in \Zb} \frac{\prod_{\mu \in \mathrm{Roots}} (q^{1 + m \cdot \mu - r/2} a^{-1} s^{\mu}; q)_{\infty}} {\prod_{\mu \in \mathrm{Roots}} (q^{1 + m \cdot \mu} s^{\mu}; q)_{\infty}}, 
\end{align}
where $\sum_{i = 1}^3 m_{n,i} = 0$.

The half-index (\ref{bdy_E6_adj_hindexA_D}) coincides with the half-index for theory B, equivalent to a known identity for a Macdonald type sum which is listed in \cite{MR2267266}
\begin{align}
\label{bdy_E6_adj_hindexB_D}
\mathbb{II}_{D^6; N^6}^{B}
&=
\frac{
(q^{1 - r}a^{-2};q)_{\infty}
(q^{1 - 5r/2}a^{-5};q)_{\infty}
(q^{1 - 3r} a^{-6};q)_{\infty}
}
{
(q^{-r/2}a^{-1};q)_{\infty}
(q^{-4r/2}a^{-4};q)_{\infty}
(q^{-5r/2}a^{-5};q)_{\infty}
}
\nonumber\\
&\times 
\frac{
(q^{1 - 4r}a^{-8};q)_{\infty}
(q^{1 - 9r/2}a^{-9};q)_{\infty}
(q^{1 - 6r}a^{-12};q)_{\infty}
}{
(q^{-7r/2}a^{-7};q)_{\infty}
(q^{-4r}a^{-8};q)_{\infty}
(q^{-11r/2}a^{-11};q)_{\infty}
}. 
\end{align}

%%%%%%%%%%%%%%%%%%%%%%%%%%%%%%%%%%
%%%%%%%%%%%%%%%%%%%%%%%%%%%%%%%%%%

%%%%%%%%%%%%%%%%%%%%%%%%%%%%%%%%%%
%%%%%%%%%%%%%%%%%%%%%%%%%%%%%%%%%%
\section{Gauge group $E_7$}
\label{sec_E7}
%%%%%%%%%%%%%%%%%%%%%%%%%%%%%%%%%%
%%%%%%%%%%%%%%%%%%%%%%%%%%%%%%%%%%
In this section we consider boundary confining dualities for $E_7$ gauge theories with fundamental or adjoint chirals. The result here for the cases with $3$ fundamental chirals is a new boundary duality corresponding to a known bulk duality. The conjectured matching of half-indices gives a new example of a generalized Askey-Wilson identity for $E_7$ which has not been studied in the mathematical literature. We give some numerical evidence for these conjectured boundary confining dualities -- see appendix~\ref{MC_int} for details.

%%%%%%%%%%%%%%%%%%%%%%%%%%%%%%%%%%
%%%%%%%%%%%%%%%%%%%%%%%%%%%%%%%%%%
\subsection{$E_7$ with $3$ fundamental chirals}
\label{sec_E7_3_integral}
%%%%%%%%%%%%%%%%%%%%%%%%%%%%%%%%%%
%%%%%%%%%%%%%%%%%%%%%%%%%%%%%%%%%%

We have dual confining boundary conditions for $E_7$ gauge theory with $3$ fundamental chirals
\begin{align}
\label{E7_3_charges}
\begin{array}{c|c|c|c|c|c}
& \textrm{bc} & E_7 & SU(N_f = 3) & U(1)_a & U(1)_R \\ \hline
\textrm{VM} & \mathcal{N} & {\bf Adj} & {\bf 1} & 0 & 0 \\
Q_{\alpha} & \textrm{N} & {\bf 56} & {\bf 3} & 1 & 0 \\
 \hline
M^{\alpha} & \textrm{N} & {\bf 1} & {\bf \overline{3}} & 2 & 0 \\
B_{\alpha \beta \gamma \delta} & \textrm{N} & {\bf 1} & {\bf 15} & 4 & 0 \\
B^{\alpha \beta \gamma} & \textrm{N} & {\bf 1} & {\bf \overline{10}} & 6 & 0 \\
B_{\alpha \beta} & \textrm{N} & {\bf 1} & {\bf 6} & 8 & 0 \\
B & \textrm{N} & {\bf 1} & {\bf 1} & 12 & 0 \\
\widehat{B} & \textrm{N} & {\bf 1} & {\bf 1} & 18 & 0 \\
V & \textrm{D} & {\bf 1} & {\bf 1} & -36 & 2
\end{array}
\end{align}

We see that the anomaly polynomials match for theories A and B:
\begin{align}
\label{bdy_E7_3_anom_A}
\Acal^A = & \underbrace{18\Tr(s^2) + \frac{133}{2}r^2}_{\textrm{VM}, \; \Ncal}
 - \underbrace{\left( 18\Tr(s^2) + 28 \Tr(x^2) + 84(a-r)^2 \right)}_{Q_{\alpha}, \; N}
   \nonumber \\
  = & - 28 \Tr(x^2) - 84a^2 + 168ar - \frac{35}{2}r^2, \\
\label{bdy_E7_3_anom_B}
\Acal^B = & - \underbrace{\left( \frac{1}{2}\Tr(x^2) + \frac{3}{2}(2a-r)^2 \right)}_{M^{\alpha}, \; N}
 - \underbrace{\left( \frac{35}{2} \Tr(x^2) + \frac{15}{2}(4a-r)^2 \right)}_{B_{\alpha \beta \gamma \delta}, \; N}
  \nonumber \\
  & - \underbrace{\left( \frac{15}{2} \Tr(x^2) + 5(6a-r)^2 \right)}_{B^{\alpha \beta \gamma}, \; N}
 - \underbrace{\left( \frac{5}{2} \Tr(x^2) + 3(8a-r)^2 \right)}_{B_{\alpha \beta}, \; N}
  \nonumber \\
  & - \underbrace{\frac{1}{2}\big( 12a - r \big)^2}_{B, \; N}
 - \underbrace{\frac{1}{2}\big( 18a - r \big)^2}_{\widehat{B}, \; N}
 + \underbrace{\frac{1}{2}\big( -36a + r \big)^2}_{V, \; D}
  \nonumber \\
  = & - 28 \Tr(x^2) - 84a^2 + 168ar - \frac{35}{2}r^2. 
\end{align}

Similarly the half-indices match. Theory A has half-index
\begin{align}
    \II^A_{\Ncal, N} = & \frac{(q)_{\infty}^7}{2^{10} \cdot 3^4 \cdot 5 \cdot 7} \prod_{i = 1}^7 \oint \frac{ds_i}{2\pi i s_i} \frac{\prod_{\mu \in \mathrm{Roots}} (s^{\mu}; q)_{\infty}}{\prod_{\alpha = 1}^3 \prod_{\mu \in \mathrm{Fund. \; weights}} (q^{r/2} s^{\mu} a x_{\alpha}; q)_{\infty}}, 
\end{align}
where $\prod_{\alpha = 1}^3 x_{\alpha} = 1$ and
\begin{align}
    \prod_{\mu \in \mathrm{Roots}} (s^{\mu}; q)_{\infty} = & \left( \prod_{1 \le i < j < k < l \le 7} (s_i^{\pm} s_j^{\pm} s_k^{\pm} s_l^{\pm}; q)_{\infty} \right) \prod_{i < j}^8 (s^{\pm}_i s^{\mp}_j; q)_{\infty}, \\
    \prod_{\mu \in \mathrm{Fund. \; weights}} (s^{\mu}; q)_{\infty} = & \prod_{i < j}^8 (s^{\pm}_i s^{\pm}_j; q)_{\infty}, 
\end{align}
with $\prod_{i = 1}^8 s_i = 1$.

Theory B has half-index
\begin{align}
    \II^B_{N^6; D} = & \frac{(q^{18r} a^{36};q)_{\infty}}{(q^{6r} a^{12};q)_{\infty} (q^{9r} a^{18};q)_{\infty} \left( \prod_{\alpha = 1}^3 ((q^{r} a^2 x_{\alpha}^{-1}; q)_{\infty} \right) \prod_{\alpha \le \beta \le \gamma \le \delta}^3 ((q^{2r} a^4 x_{\alpha} x_{\beta} x_{\gamma} x_{\delta};q)_{\infty}}
    \nonumber \\
     & \times \frac{1}{\left( \prod_{\alpha \le \beta \le \gamma}^3 ((q^{3r} a^6 x_{\alpha}^{-1} x_{\beta}^{-1} x_{\gamma}^{-1}; q)_{\infty} \right) \prod_{\alpha \le \beta}^3 ((q^{4r} a^8 x_{\alpha} x_{\beta};q)_{\infty}}. 
\end{align}

As for the cases with $E_7$ gauge group, calculating the $q$-series expansion of the theory A half-index is computationally expensive. However, we can consider the reduced half-index given by taking the $q \to 0$ limit after fixing $\mathfrak{t} = q^{r/2} a$ and $x_{\alpha}$. We have checked that the Mathematica Monte Carlo numerical integration of the theory A reduced half-index matches the evaluation of the theory B reduced half-index for various choices of $\mathfrak{t}$ and $x_{\alpha}$ such that $|\mathfrak{t} x_{\alpha}| < 1$ to the order of approximately $1\%$, providing strong numerical evidence for the claimed boundary confining duality. Details are given in appendix~\ref{MC_int}.

With the opposite boundary conditions we have
\begin{align}
    \II^A_{\Dcal, D} = & \frac{1}{(q)_{\infty}^7} \sum_{m_i \in \Zb} \frac{\prod_{\alpha = 1}^3 \prod_{\mu \in \mathrm{Fund. \; weights}} (q^{1 + m \cdot \mu - r/2} s^{\mu} a^{-1} x_{\alpha}^{-1}; q)_{\infty}} {\prod_{\mu \in \mathrm{Roots}} (q^{1 + m \cdot \mu} s^{\mu}; q)_{\infty}}, 
\end{align}
with $\sum_{i = 1}^8 m_i = 0$.

Theory B has matching half-index
\begin{align}
    \II^B_{D^6; N} = & \frac{(q^{1 - 6r} a^{-12};q)_{\infty} (q^{1 - 9r} a^{-18};q)_{\infty} \prod_{\alpha \le \beta \le \gamma \le \delta}^3 (q^{1 - 2r} a^{-4} x_{\alpha}^{-1} x_{\beta}^{-1} x_{\gamma}^{-1} x_{\delta}^{-1};q)_{\infty}} {(q^{1 - 18r} a^{-36};q)_{\infty}}
    \nonumber \\
     & \times \left( \prod_{\alpha = 1}^3 (q^{1 - r} a^{-2} x_{\alpha}; q)_{\infty} \right) \left( \prod_{\alpha \le \beta \le \gamma}^3 (q^{1 - 3r} a^{-6} x_{\alpha} x_{\beta} x_{\gamma}; q)_{\infty} \right) \prod_{\alpha \le \beta}^3 (q^{1 - 4r} a^{-8} x_{\alpha}^{-1} x_{\beta}^{-1};q)_{\infty}. 
\end{align}
We conjecture a new Macdonald type sum identity corresponding to the matching of these half-indices.

%%%%%%%%%%%%%%%%%%%%%%%%%%%%%%%%%%
%%%%%%%%%%%%%%%%%%%%%%%%%%%%%%%%%%
\subsection{$E_7$ with an adjoint chiral}
\label{sec_E7_adj_integral}
%%%%%%%%%%%%%%%%%%%%%%%%%%%%%%%%%%
%%%%%%%%%%%%%%%%%%%%%%%%%%%%%%%%%%

The charges and boundary conditions are summarized as follows: 
\begin{align}
\label{E7_adj_charges}
\begin{array}{c|c|c|c|c}
& \textrm{bc} & E_7 & U(1)_a & U(1)_R \\ \hline
\textrm{VM} & \mathcal{N} & {\bf Adj} & 0 & 0 \\
\Phi & \textrm{N} & {\bf Adj} & 1 & 0 \\
 \hline
M_I & \textrm{N} & {\bf 1} & I & 0 \\
V_{I-1} & \textrm{D} & {\bf 1} & -(I-1) & 0
\end{array}
\end{align}
where $I \in \{2, 6, 8, 10, 12, 14, 18\}$.

The boundary 't Hooft anomaly of theory A is given by
\begin{align}
\label{bdy_E7_adj_anomA}
\Acal ^A& = \underbrace{{18\Tr(s^2)} + \frac{133}{2}r^2}_{\textrm{VM}, \; \Ncal}
 - \underbrace{\left( 18\Tr(s^2) + \frac{133}{2}(a-r)^2 \right)}_{\Phi, \; N}
  \nonumber \\
  = & - \frac{133}{2}a^2 + 133ar, 
\end{align}
which exactly agrees with the boundary anomaly of theory B which is given by
\begin{align}
\label{bdy_E7_adj_anomB}
\Acal ^B= & \sum_{I \in \{2, 6, 8, 10, 12, 14, 18\}} \left(
   - \underbrace{\frac{1}{2}\big(Ia - r\big)^2}_{M_I, \; N}
   + \underbrace{\frac{1}{2}\big((1-I)a - r\big)^2}_{V_{I-1}, \; D} \right). 
\end{align}

This gives the half-index for theory A
\begin{align}
\label{bdy_E7_adj_hindexA}
    \II^A_{\Ncal, N} = & \frac{1}{2^{10} \cdot 3^4 \cdot 5 \cdot 7}
    \frac{(q)_{\infty}^7}{(q^{r/2} a; q)_{\infty}^7}
    \prod_{i = 1}^7 \oint \frac{ds_i}{2\pi i s_i} \frac{\prod_{\mu \in \mathrm{Roots}} (s^{\mu}; q)_{\infty}}{\prod_{\mu \in \mathrm{Roots}} (q^{r/2} a s^{\mu}; q)_{\infty}}, 
\end{align}
where
\begin{align}
    \prod_{\mu \in \mathrm{Roots}} (s^{\mu}; q)_{\infty} = & \left( \prod_{1 \le i < j < k < l \le 7} (s_i^{\pm} s_j^{\pm} s_k^{\pm} s_l^{\pm}; q)_{\infty} \right) \prod_{i < j}^8 (s^{\pm}_i s^{\mp}, _j; q)_{\infty}, 
\end{align}
with $\prod_{i = 1}^8 s_i = 1$.

The half-index (\ref{bdy_E7_adj_hindexA}) coincides \cite{MR1314036} with the half-index for theory B
\begin{align}
\label{bdy_E7_adj_hindexB}
\mathbb{II}_{N^7; D^7}^{B}
&=
\frac{
(q^{1+r/2}a;q)_{\infty}
(q^{1+5r/2}a^5;q)_{\infty}
(q^{1+7r/2}a^7;q)_{\infty}
(q^{1+9r/2}a^9;q)_{\infty}
}
{
(q^{r}a^{2};q)_{\infty}
(q^{3r}a^{6};q)_{\infty}
(q^{4r} a^{8};q)_{\infty}
(q^{5r} a^{10};q)_{\infty}
}
\nonumber\\
&\times 
\frac{
(q^{1+11r/2}a^{11};q)_{\infty}
(q^{1+13r/2}a^{13};q)_{\infty}
(q^{1+17r/2}a^{17};q)_{\infty}
}
{
(q^{6r}a^{12};q)_{\infty}
(q^{7r}a^{14};q)_{\infty}
(q^{9r}a^{18};q)_{\infty}
}. 
\end{align}

With the opposite boundary conditions we have
\begin{align}
\label{bdy_E7_adj_hindexA_D}
    \II^A_{\Dcal, D} = & 
    \frac{(q^{1 - r/2} a^{-1}; q)_{\infty}^7} {(q)_{\infty}^7}
    \sum_{m_i \in \Zb} \frac{\prod_{\mu \in \mathrm{Roots}} (q^{1 + m \cdot \mu - r/2} a^{-1} s^{\mu}; q)_{\infty}} {\prod_{\mu \in \mathrm{Roots}} (q^{1 + m \cdot \mu} s^{\mu}; q)_{\infty}}, 
\end{align}
where $\sum_{i = 1}^8 m_i = 0$.

The half-index (\ref{bdy_E7_adj_hindexA_D}) coincides with the half-index for theory B, equivalent to a known identity for a Macdonald type sum which is listed in \cite{MR2267266}
\begin{align}
\label{bdy_E7_adj_hindexB_D}
\mathbb{II}_{D^7; N^7}^{B}
&=
\frac{
(q^{1 - r}a^{-2};q)_{\infty}
(q^{1 - 3r}a^{-6};q)_{\infty}
(q^{1 - 4r} a^{-8};q)_{\infty}
(q^{1 - 5r} a^{-10};q)_{\infty}
}
{
(q^{-r/2}a^{-1};q)_{\infty}
(q^{-5r/2}a^{-5};q)_{\infty}
(q^{-7r/2}a^{-7};q)_{\infty}
(q^{-9r/2}a^{-9};q)_{\infty}
}
\nonumber\\
&\times 
\frac{
(q^{1 - 6r}a^{-12};q)_{\infty}
(q^{1 - 7r}a^{-14};q)_{\infty}
(q^{1 - 9r}a^{-18};q)_{\infty}
}
{
(q^{-11r/2}a^{-11};q)_{\infty}
(q^{-13r/2}a^{-13};q)_{\infty}
(q^{-17r/2}a^{-17};q)_{\infty}
}. 
\end{align}

%%%%%%%%%%%%%%%%%%%%%%%%%%%%%%%%%%
%%%%%%%%%%%%%%%%%%%%%%%%%%%%%%%%%%

%%%%%%%%%%%%%%%%%%%%%%%%%%%%%%%%%%
%%%%%%%%%%%%%%%%%%%%%%%%%%%%%%%%%%
\section{$E_8$ with an adjoint chiral}
\label{sec_E8_adj_integral}
%%%%%%%%%%%%%%%%%%%%%%%%%%%%%%%%%%
%%%%%%%%%%%%%%%%%%%%%%%%%%%%%%%%%%
For $E_8$ there is no distinction between the fundamental and adjoint representations so we have only one case to consider for a boundary confining duality. It is most naturally interpreted as an example with an adjoint chiral.

The charges and boundary conditions are summarized as follows: 
\begin{align}
\label{E8_adj_charges}
\begin{array}{c|c|c|c|c}
& \textrm{bc} & E_8 & U(1)_a & U(1)_R \\ \hline
\textrm{VM} & \mathcal{N} & {\bf Adj} & 0 & 0 \\
\Phi & \textrm{N} & {\bf Adj} & 1 & 0 \\
 \hline
M_I & \textrm{N} & {\bf 1} & I & 0 \\
V_{I-1} & \textrm{D} & {\bf 1} & -(I-1) & 0
\end{array}
\end{align}
where $I \in \{2, 8, 12, 14, 18, 20, 24, 30\}$.

The boundary 't Hooft anomaly of theory A is given by
\begin{align}
\label{bdy_E8_adj_anomA}
\Acal ^A& = \underbrace{{30\Tr(s^2)} + 124r^2}_{\textrm{VM}, \; \Ncal}
 - \underbrace{\left( 30\Tr(s^2) +124(a-r)^2 \right)}_{\Phi, \; N}
  \nonumber \\
  = & - 124a^2 + 248ar , 
\end{align}
which exactly agrees with the boundary anomaly of theory B which is given by
\begin{align}
\label{bdy_E8_adj_anomB}
\Acal ^B= & \sum_{I \in \{2, 8, 12, 14, 18, 20, 24, 30\}} \left(
   - \underbrace{\frac{1}{2}\big(Ia - r\big)^2}_{M_I, \; N}
   + \underbrace{\frac{1}{2}\big((1-I)a - r\big)^2}_{V_{I-1}, \; D} \right) . 
\end{align}

This gives the half-index for theory A
\begin{align}
\label{bdy_E8_adj_hindexA}
\mathbb{II}_{\mathcal{N},N}^{A}
&=
\frac{1}{2^{14}\cdot 3^{5}\cdot 5^2\cdot 7}
\frac{(q)_{\infty}^8}
{(q^{r/2}a;q)_{\infty}^{8}}
\prod_{i=1}^{8}
\oint \frac{ds_{i}}{2\pi is_{i}}
\nonumber\\
&\times 
\prod_{1\le i<j\le8}
\frac{
\left( (s_{i}^2 s_{j}^2 \prod_{n=1}^{8}s_{n}^{-1})^{\pm};q \right)_{\infty}
}
{
\left(q^{r/2}a (s_{i}^2 s_{j}^2 \prod_{n=1}^{8}s_{n}^{-1})^{\pm};q \right)_{\infty}
}
\frac{
\left( (\prod_{n=1}^{8}s_{n})^{\pm};q \right)_{\infty}
}
{
\left( q^{r/2}a (\prod_{n=1}^{8}s_{n})^{\pm} \right)_{\infty}
}
\nonumber\\
&\times 
\prod_{1\le i<j\le8}
\frac{
(s_{i}^2 s_{j}^{\pm2};q)_{\infty}
(s_{i}^{-2}s_{j}^{\pm2};q)_{\infty}
}
{
(q^{r/2} a s_{i}^2 s_{j}^{\pm2};q)_{\infty}
(q^{r/2} a s_{i}^{-2}s_{j}^{\pm2};q)_{\infty}
}
\nonumber\\
&\times 
\prod_{1\le i<j<l<m\le8} 
\frac{
\left( (s_{i}^2 s_{j}^2 s_{l}^2 s_{m}^2 \prod_{n=1}^{8}s_{n}^{-1})^{\pm};q \right)_{\infty}
}
{
\left(q^{r/2}a (s_{i}^2 s_{j}^2 s_{l}^2 s_{m}^2 \prod_{n=1}^{8}s_{n}^{-1})^{\pm};q \right)_{\infty}
}. 
\end{align}

The half-index (\ref{bdy_E8_adj_hindexA}) coincides \cite{MR1314036} with the half-index for theory B
\begin{align}
\label{bdy_E8_adj_hindexB}
\mathbb{II}_{N^8;D^8}^{B}
&=
\frac{
(q^{1+r/2}a;q)_{\infty}
(q^{1+7r/2}a^7;q)_{\infty}
(q^{1+11/r}a^{11};q)_{\infty}
(q^{1+13r/2} a^{13};q)_{\infty}
}
{
(q^{r}a^{2};q)_{\infty}
(q^{4r}a^{8};q)_{\infty}
(q^{6r} a^{12};q)_{\infty}
(q^{7r} a^{14};q)_{\infty}
}
\nonumber\\
&\times 
\frac{
(q^{1+17r/2}a^{17};q)_{\infty}
(q^{1+19r/2}a^{19};q)_{\infty}
(q^{1+23r/2}a^{23};q)_{\infty}
(q^{1+29r/2}a^{29};q)_{\infty}
}
{
(q^{9r}a^{18};q)_{\infty}
(q^{10r}a^{20};q)_{\infty}
(q^{12r}a^{24};q)_{\infty}
(q^{15r}a^{30};q)_{\infty}
}. 
\end{align}

Switching all boundary conditions gives
\begin{align}
\label{bdy_E8_adj_hindexA_D}
\mathbb{II}_{\mathcal{D},D}^{A}
&=
\frac{(q^{1 - r/2}a^{-1};q)_{\infty}^{8}}
{(q)_{\infty}^8}
\sum_{m_i \in \Zb}
\nonumber\\
&\times 
\prod_{1\le i<j\le8}
\frac{
\left(q^{1 - r/2}a^{-1} (\tilde{s}_{i}^2 \tilde{s}_{j}^2 \prod_{n=1}^{8}\tilde{s}_{n}^{-1})^{\pm};q \right)_{\infty}
}
{
\left( (\tilde{s}_{i}^2 \tilde{s}_{j}^2 \prod_{n=1}^{8}\tilde{s}_{n}^{-1})^{\pm};q \right)_{\infty}
}
\frac{
\left( q^{1 - r/2}a^{-1} (\prod_{n=1}^{8}\tilde{s}_{n})^{\pm} \right)_{\infty}
}
{
\left( (\prod_{n=1}^{8}\tilde{s}_{n})^{\pm};q \right)_{\infty}
}
\nonumber\\
&\times 
\prod_{1\le i<j\le8}
\frac{
(q^{1 - r/2} a^{-1} \tilde{s}_{i}^2 \tilde{s}_{j}^{\pm2};q)_{\infty}
(q^{1 - r/2} a^{-1} \tilde{s}_{i}^{-2}\tilde{s}_{j}^{\pm2};q)_{\infty}
}
{
(\tilde{s}_{i}^2 \tilde{s}_{j}^{\pm2};q)_{\infty}
(\tilde{s}_{i}^{-2}\tilde{s}_{j}^{\pm2};q)_{\infty}
}
\nonumber\\
&\times 
\prod_{1\le i<j<l<m\le8} 
\frac{
\left(q^{1 - r/2}a^{-1} (\tilde{s}_{i}^2 \tilde{s}_{j}^2 \tilde{s}_{l}^2 \tilde{s}_{m}^2 \prod_{n=1}^{8}\tilde{s}_{n}^{-1})^{\pm};q \right)_{\infty}
}
{
\left( (\tilde{s}_{i}^2 \tilde{s}_{j}^2 \tilde{s}_{l}^2 \tilde{s}_{m}^2 \prod_{n=1}^{8}\tilde{s}_{n}^{-1})^{\pm};q \right)_{\infty}
}, 
\end{align}
where $\tilde{s}_i \equiv q^{m_i} s_i$.

The half-index (\ref{bdy_E8_adj_hindexA_D}) coincides with the half-index for theory B, equivalent to a known identity for a Macdonald type sum which is listed in \cite{MR2267266}
\begin{align}
\label{bdy_E8_adj_hindexB_D}
\mathbb{II}_{D^8;N^8}^{B}
&=
\frac{
(q^{1 - r}a^{-2};q)_{\infty}
(q^{1 - 4r}a^{-8};q)_{\infty}
(q^{1 - 6r} a^{-12};q)_{\infty}
(q^{1 - 7r} a^{-14};q)_{\infty}
}
{
(q^{-r/2}a^{-1};q)_{\infty}
(q^{-7r/2}a^{-7};q)_{\infty}
(q^{-11/r}a^{-11};q)_{\infty}
(q^{-13r/2} a^{-13};q)_{\infty}
}
\nonumber\\
&\times 
\frac{
(q^{1 - 9r}a^{-18};q)_{\infty}
(q^{1 - 10r}a^{-20};q)_{\infty}
(q^{1 - 12r}a^{-24};q)_{\infty}
(q^{1 - 15r}a^{-30};q)_{\infty}
}
{
(q^{-17r/2}a^{-17};q)_{\infty}
(q^{-19r/2}a^{-19};q)_{\infty}
(q^{-23r/2}a^{-23};q)_{\infty}
(q^{-29r/2}a^{-29};q)_{\infty}
}. 
\end{align}

%%%%%%%%%%%%%%%%%%%%%%%%%%%%%%%%%%%
\appendix
%%%%%%%%%%%%%%%%%%%%%%%%%%%%%%%%%%%
\section{Numerical checks of $E_6$ and $E_7$ refined Hilbert series identities}
\label{MC_int}
%%%%%%%%%%%%%%%%%%%%%%%%%%%%%%%%%%%

We present some tables of numerical evaluations of $\rII^A$ using Mathematica NIntegrate with Monte Carlo method and $10^{11}$ samples in the cases of $E_6$ with $N_f + N_a = 4$ and $E_7$ with $N_f = 3$. These give evidence for the matching of the refined reduced half-indices (boundary Hilbert series) and hence for the boundary confining dualities.

The flavour fugacities are chosen randomly within the unit circle. We present $10$ such examples for each case theory, with the first table giving the fugacity values and the second showing the ration of the evaluated reduced half-indices for theories A and B. Note that the reduced half-indices in each case typically match to better than 1\%. However, the theory A integrals are numerically challenging with lots of cancellations, so some cases the relative numerical error can be much larger than 1\%. The theory B reduced half-indices are rational functions so will be evaluated exactly to the number of significant figures presented. For comparison we also list equivalent numerical integration results for the case of $G_2$ with $N_f = 4$. In that case, the matching is a known identity but we see similar accuracy in numerical comparison (even though this is only a $2$-dimensional integral).

\subsection{$E_6$ -- $[4]$}

\small
$$
\begin{array}{c|c}
 & \mathfrak{t} x_1, \mathfrak{t} x_2, \mathfrak{t} x_3, \mathfrak{t} x_4
 \\ \hline
1 & -0.0234759+0.00711192 i, 0.496534 +0.610979 i, -0.326678+0.169003 i, 0.0828884 +0.68146 i

\\

2 & 0.302088 +0.313318 i, 0.0140813 -0.0538188 i, -0.163671-0.199149 i, 0.0451775 -0.15439 i

\\

3 & -0.0250033+0.166172 i, -0.291018-0.37288 i, 0.183273 -0.0258147 i, -0.0495631+0.0706467 i

\\

4 & 0.685339 -0.129819 i, 0.50425 -0.326404 i, -0.24908-0.133475 i, -0.0947395+0.0335816 i

\\

5 & 0.190953 -0.208135 i, 0.239949 -0.0329212 i, 0.137727 -0.355568 i, 0.109013 -0.151154 i

\\

6 & -0.170367-0.0648944 i, 0.18149 +0.384322 i, 0.0712975 +0.0470663 i, 0.321017 +0.545043 i

\\

7 & -0.455215+0.310856 i, -0.01625-0.00814275 i, -0.0535401+0.994244 i, 0.361642 +0.158975 i

\\ 

8 & 0.23476 +0.11059 i, 0.40632 +0.228978 i, -0.725015-0.0481284 i, -0.692562+0.272425 i

\\ 

9 & 0.275143 +0.0751123 i, 0.30426 -0.00926237 i, 0.439344 +0.111412 i, -0.591255+0.754494 i

\\ 

10 & -0.0521672-0.113462 i, 0.657591 -0.366998 i, 0.896003 +0.358691 i, -0.763347-0.0959539 i

\end{array}
$$

$$
\begin{array}{c|c|c}
 & \rII^B & \rII^A/\rII^B
 \\ \hline
1

& 0.290701 -0.299555 i

& 1.03254 +0.0123426 i

\\

2

& 0.991963 +0.0363137 i

& 0.991567 +0.00135808 i

\\

3

& 1.06149 -0.0384057 i

& 0.993826 -0.00103762 i

\\

4

& 0.826482 -0.858645 i

& 1.00281 -0.00120006 i

\\

5

& 0.760051 -0.113749 i

& 0.988162 -0.00166513 i

\\

6

& 0.598655 -0.0512807 i

& 1.00783 +0.00417579 i

\\

7

& 0.348527 -1.11526 i

& 1.02099 -0.324376 i

\\ 

8

& 0.645306 +0.400197 i

& 0.993663 +0.00940664 i

\\ 
9

& 1.29865 +1.01025 i

& 0.987966 -0.0289081 i

\\ 

10

& 2.72067 +1.40683 i

& 1.01216 -0.00880855 i
\end{array}
$$
\normalsize

\subsection{$E_6$ -- $[3+1]$}

\small
$$
\begin{array}{c|c}
 & \mathfrak{t} x_1, \mathfrak{t} x_2, \mathfrak{t} x_3, \tilde{\mathfrak{t}}
 \\ \hline
1 &
-0.751832+0.00396913 i, -0.0971231+0.00130853 i, -0.284908+0.102964 i, -0.110992+0.990821 i

\\ 
2 &
-0.317032+0.779203 i, -0.140079-0.89742 i, 0.00280996 -0.00357397 i, 0.558137 +0.012314 i

\\ 
3 &
-0.190562+0.0957396 i, -0.355162-0.362745 i, -0.150115+0.896124 i, -0.323203-0.557585 i

\\ 
4 &
0.0464883 -0.177868 i, 0.00293396 +0.5284 i, -0.502174-0.853927 i, -0.207428+0.43438 i

\\ 
5 &
-0.00129523-0.000809934 i, 0.086667 -0.417337 i, -0.29446+0.303761 i, 0.122194 -0.266339 i

\\ 
6 &
0.113931 -0.362205 i, -0.08484-0.203071 i, -0.256592+0.560824 i, -0.695269+0.511302 i

\\ 
7 &
-0.531755-0.47155 i, 0.417641 +0.832083 i, -0.0743604+0.217454 i, -0.111881-0.181178 i

\\ 
8 &
-0.00951035+0.0135762 i, 0.0573461 -0.273941 i, -0.339735-0.374335 i, -0.255882-0.102164 i

\\ 
9 &
-0.784786+0.338097 i, -0.229029-0.388009 i, -0.31641-0.310386 i, -0.395246+0.485849 i

\\ 
10 &
0.0938797 +0.406379 i, -0.522526+0.706919 i, 0.243731 -0.499185 i, 0.655442 +0.401781 i

\end{array}
$$

$$
\begin{array}{c|c|c}
  & \rII^B & \rII^A/\rII^B
 \\ \hline

1

& -0.0367995-0.151403 i

& 1.09724 -0.120556 i

\\ 

2

& 1.1731 +0.204169 i

& 1.00162 -0.00807585 i

\\ 

3

& 2.95137 -1.8163 i

& 0.990775 +0.0110375 i

\\ 

4

& 42.7386 -32.547 i

& 1.0001 -0.0084357 i

\\ 

5

& 0.923462 +0.137879 i

& 0.996371 -0.00324904 i

\\ 

6

& 0.773316 +0.598055 i

& 0.987898 -0.0126232 i

\\ 

7

& 0.959142 -0.66293 i

& 0.993842 -0.00968543 i

\\ 

8

& 1.12691 +0.196216 i

& 0.985935 +0.00401361 i

\\ 

9

& 0.400957 -0.901223 i

& 1.03522 +0.0286261 i

\\ 

10

& 1.58318 +0.259849 i

& 0.996495 +0.00061966 i
\end{array}
$$
\normalsize

\subsection{$E_6$ -- $[2+2]$}

\small
$$
\begin{array}{c|c}
& \mathfrak{t} x_1, \mathfrak{t} x_2, \tilde{\mathfrak{t}} \tilde{x}_1, \tilde{\mathfrak{t}} \tilde{x}_2
 \\ \hline
1 &
0.640038 +0.53923 i, -0.268455-0.84344 i, 0.386667 -0.806928 i, 0.010295 -0.207447 i

\\ 
2 &
0.737773 +0.2534 i, -0.16533-0.232279 i, -0.38595+0.775523 i, -0.623401-0.20632 i

\\ 
3 &
0.109171 -0.585594 i, -0.793789+0.0964344 i, 0.279103 -0.134808 i, 0.0354895 +0.380692 i

\\ 
4 &
0.7167 -0.383209 i, 0.101425 +0.389796 i, 0.0839903 +0.0738067 i, -0.037217-0.0324649 i

\\ 
5 &
0.0230923 -0.0765744 i, 0.585828 +0.258123 i, 0.347055 -0.479408 i, -0.160716+0.578402 i

\\ 
6 &
0.067434 -0.318034 i, 0.00650358 +0.00587156 i, 0.00240265 +0.0439869 i, -0.0368151+0.102164 i

\\ 
7 &
0.030406 +0.000694502 i, -0.159088-0.106352 i, 0.6817 +0.0870333 i, -0.0849712+0.326787 i

\\ 
8 &
-0.448639-0.191149 i, -0.604296+0.25106 i, -0.84651-0.456231 i, -0.418166+0.446437 i

\\ 
9 &
0.0454495 -0.577359 i, 0.495194 +0.000122798 i, -0.26322+0.104288 i, 0.0123185 -0.0233005 i

\\ 
10 &
-0.360204-0.362996 i, 0.658296 -0.624635 i, 0.272641 +0.195759 i, 0.0429119 -0.0867349 i

\end{array}
$$

$$
\begin{array}{c|c|c}
 & \rII^B & \rII^A/\rII^B
 \\ \hline
1 &

1.15761 -0.17734 i

& 1.02376 +0.0135868 i

\\ 
2 &

1.65281 +0.416576 i

& 0.991742 -0.00349281 i

\\ 
3 &

0.589971 -0.228187 i

& 0.974194 -0.0293323 i

\\ 
4 &

1.07735 -0.326778 i

& 1.00913 -0.00814665 i

\\ 
5 &

1.1857 -0.046149 i

& 1.00832 -0.00136545 i

\\ 
6 &

1.02442 +0.049962 i

& 1.00376 +0.00238796 i

\\ 
7 &

1.19983 +0.15792 i

& 0.996318 -0.00325625 i

\\ 
8 &

2.86869 +0.260658 i

& 1.00807 +0.00773062 i

\\ 
9 &

0.814242 +0.166933 i

& 1.00164 +0.00296507 i

\\ 
10 &

0.788741 -0.284616 i

& 1.00576 +0.00282165 i
\end{array}
$$
\normalsize

\subsection{$E_7$ -- $[3]$}

\small
$$
\begin{array}{c|c}
 & \mathfrak{t} x_1, \mathfrak{t} x_2, \mathfrak{t} x_3
 \\ \hline
 1 &
 -0.212731+0.922799 i, 0.114579 +0.268194 i, 0.0614174 -0.0300152 i

\\
2 &
  0.0835584 +0.223947 i, 0.302454 +0.111834 i, -0.391574-0.0772544 i

  \\
  3 &
-0.376041-0.245701 i, -0.250929+0.260853 i, 0.626473 +0.249931 i

\\
4 &
0.00383335 -0.0756743 i, -0.782634+0.251244 i, 0.108969 +0.021749 i

\\
5 &
0.00383962 -0.00433834 i, 0.470105 -0.391208 i, -0.500468+0.101544 i

\\
6 &
0.00725631 +0.242431 i, -0.000621355-0.587519 i, -0.0760336+0.0527554 i

\\
7 &
-0.0529269+0.0145817 i, 0.174268 +0.0158707 i, 0.4198 -0.0719301 i

\\
8 &
-0.157932-0.0235142 i, -0.234763-0.797787 i, 0.0424514 +0.00463794 i

\\
9 &
-0.0676571-0.620083 i, 0.824613 +0.0368542 i, 0.496211 -0.134759 i

\\
10 &
0.0663995 -0.017729 i, -0.402693+0.302654 i, -0.165837+0.26069 i

\end{array}
$$

$$
\begin{array}{c|c|c}
 & \rII^B & \rII^A/\rII^B
 \\ \hline
1 & 0.768066 +1.16539 i
  & 1.0144 -0.0139891 i
\\
2 & 0.882036 -0.0624183 i
  & 1.00039 -0.000403221 i
  \\
3 & 0.765711 -0.0555455 i
&0.977582 +0.00160989 i
\\
4 & 0.907271 -0.325205 i
& 1.00847 -0.000441181 i
\\
5 & 0.78249 +0.156931 i
& 1.00442 -0.00500289 i
\\
6 & 1.2972 +0.0197836 i
& 1.00008 +0.000310746 i
\\
7 & 1.09277 -0.0202721 i
& 1.00913 -0.000147894 i
\\
8 & 0.930522 -0.441435 i
& 0.996598 -0.00105783 i
\\
9 & 1.26266 -2.09384 i
& 0.99619 +0.0000279944 i
\\
10 & 0.837406 -0.0978072 i
& 1.00388 -0.00224307 i
\end{array}
$$
\normalsize

\subsection{$G_2$ -- $[4]$}

\small
$$
\begin{array}{c|c}
 & \mathfrak{t} x_1, \mathfrak{t} x_2, \mathfrak{t} x_3, \mathfrak{t} x_4
 \\ \hline
 1 &
 0.0481169 -0.00954673 i, -0.348491+0.0943523 i, -0.0316049+0.146036 i, -0.241138-0.160531 i

\\
2 &
  0.0426109 -0.271141 i, 0.00864091 +0.344733 i, 0.271189 -0.200707 i, 0.153095 -0.0210072 i

  \\
  3 &
0.100328 +0.216854 i, 0.197554 -0.169338 i, -0.202581-0.0579197 i, -0.0990293+0.527108 i

\\
4 &
0.779234 -0.378523 i, 0.215991 -0.400936 i, -0.141533+0.648036 i, 0.141146 -0.491235 i

\\
5 &
-0.195434+0.215521 i, 0.374101 -0.579846 i, 0.0373534 +0.168928 i, 0.757218 +0.271768 i

\\
6 &
0.51837 -0.202005 i, 0.154656 +0.234617 i, 0.41523 +0.475004 i, -0.870508-0.00263467 i

\\
7 &
-0.0558345+0.155365 i, 0.161043 +0.269449 i, -0.330075+0.392033 i, 0.289452 +0.23516 i

\\
8 &
0.890437 +0.40596 i, -0.0656358+0.435523 i, -0.00233228-0.0562747 i, -0.0391703-0.691975 i

\\
9 &
-0.846045+0.027989 i, 0.252581 -0.195542 i, 0.813427 +0.129812 i, -0.108949-0.076414 i

\\
10 &
0.666819 +0.0422453 i, 0.513887 -0.364365 i, 0.132615 +0.275064 i, 0.499662 -0.227566 i

\end{array}
$$

$$
\begin{array}{c|c|c}
 & \rII^B & \rII^A/\rII^B
 \\ \hline
1 & 1.25477 -0.0400717 i
  & 0.99436 +0.00017306 i
\\
2 & 1.07994 -0.159265 i
  & 0.996044 -0.00191212 i
  \\
3 & 0.783351 -0.0359792 i
& 0.999261 -0.000784243 i
\\
4 & 0.47606 -1.37787 i
& 0.985373 +0.00539493 i
\\
5 & 1.42525 +0.586382 i
& 0.985169 +0.00329354 i
\\
6 & 2.03766 +0.00186711 i
& 0.988306 +0.00382371 i
\\
7 & 0.524428 -0.0300966 i
& 0.990844 -0.0340454 i
\\
8 & 0.705707 +0.776239 i
& 0.964218 +0.0377104 i
\\
9 & 4.48897 +1.74889 i
& 0.996121 -0.000867226 i
\\
10 & 6.87446 -10.4909 i
& 1.00913 +0.00694605 i
\end{array}
$$
\normalsize

%%%%%%%%%%%%%%%%%%%%%%%%%%%%%%%%%%%
\subsection*{Acknowledgements}
The authors would like to thank Masahiko Ito and Masatoshi Noumi for useful discussions and comments. 
The work of T.O. was supported by the Startup Funding no.\ 4007012317 of the Southeast University.
The research of DJS was supported in part by the STFC Consolidated grant ST/T000708/1.
%%%%%%%%%%%%%%%%%%%%%%%%%%%%%%%%%%%

%%%%%%%%%%%%%%%%%%%%%%%%%%%%%%%%%%
%%%%%%%%%%%%%%%%%%%%%%%%%%%%%%%%%%

%%%%%%%%%%%%%%%%%%%%%%%%%%%%%%%%%%
%%%%%%%%%%%%%%%%%%%%%%%%%%%%%%%%%%
\bibliographystyle{utphys}
\bibliography{ref}

\def\polhk#1{\setbox0=\hbox{#1}{\ooalign{\hidewidth
  \lower1.5ex\hbox{`}\hidewidth\crcr\unhbox0}}} \def\cprime{$'$}
\providecommand{\href}[2]{#2}\begingroup\raggedright\begin{thebibliography}{10}

\bibitem{Intriligator:1995ne}
K.~A. Intriligator and P.~Pouliot, ``{Exact superpotentials, quantum vacua and
  duality in supersymmetric SP(N(c)) gauge theories},''
  \href{https://dx.doi.org/10.1016/0370-2693(95)00618-U}{{\em Phys. Lett. B}
  {\bfseries 353} (1995) 471--476},
  \href{https://arxiv.org/abs/hep-th/9505006}{{\ttfamily
  arXiv:hep-th/9505006}}.

\bibitem{Berkooz:1995km}
M.~Berkooz, ``{The Dual of supersymmetric SU(2k) with an antisymmetric tensor
  and composite dualities},''
  \href{https://dx.doi.org/10.1016/0550-3213(95)00400-M}{{\em Nucl. Phys. B}
  {\bfseries 452} (1995) 513--525},
  \href{https://arxiv.org/abs/hep-th/9505067}{{\ttfamily
  arXiv:hep-th/9505067}}.

\bibitem{Pouliot:1995me}
P.~Pouliot, ``{Duality in SUSY SU(N) with an antisymmetric tensor},''
  \href{https://dx.doi.org/10.1016/0370-2693(95)01427-6}{{\em Phys. Lett. B}
  {\bfseries 367} (1996) 151--156},
  \href{https://arxiv.org/abs/hep-th/9510148}{{\ttfamily
  arXiv:hep-th/9510148}}.

\bibitem{Luty:1996cg}
M.~A. Luty, M.~Schmaltz, and J.~Terning, ``{A Sequence of duals for Sp(2N)
  supersymmetric gauge theories with adjoint matter},''
  \href{https://dx.doi.org/10.1103/PhysRevD.54.7815}{{\em Phys. Rev. D}
  {\bfseries 54} (1996) 7815--7824},
  \href{https://arxiv.org/abs/hep-th/9603034}{{\ttfamily
  arXiv:hep-th/9603034}}.

\bibitem{Csaki:1996sm}
C.~Csaki, M.~Schmaltz, and W.~Skiba, ``{A Systematic approach to confinement in
  N=1 supersymmetric gauge theories},''
  \href{https://dx.doi.org/10.1103/PhysRevLett.78.799}{{\em Phys. Rev. Lett.}
  {\bfseries 78} (1997) 799--802},
  \href{https://arxiv.org/abs/hep-th/9610139}{{\ttfamily
  arXiv:hep-th/9610139}}.

\bibitem{Csaki:1996zb}
C.~Csaki, M.~Schmaltz, and W.~Skiba, ``{Confinement in N=1 SUSY gauge theories
  and model building tools},''
  \href{https://dx.doi.org/10.1103/PhysRevD.55.7840}{{\em Phys. Rev. D}
  {\bfseries 55} (1997) 7840--7858},
  \href{https://arxiv.org/abs/hep-th/9612207}{{\ttfamily
  arXiv:hep-th/9612207}}.

\bibitem{Terning:1997jj}
J.~Terning, ``{Duals for SU(N) SUSY gauge theories with an antisymmetric
  tensor: Five easy flavors},''
  \href{https://dx.doi.org/10.1016/S0370-2693(98)00074-4}{{\em Phys. Lett. B}
  {\bfseries 422} (1998) 149--157},
  \href{https://arxiv.org/abs/hep-th/9712167}{{\ttfamily
  arXiv:hep-th/9712167}}.

\bibitem{Garcia-Etxebarria:2012ypj}
I.~Garc\'\i{}a-Etxebarria, B.~Heidenreich, and T.~Wrase, ``{New N=1 dualities
  from orientifold transitions. Part I. Field Theory},''
  \href{https://dx.doi.org/10.1007/JHEP10(2013)007}{{\em JHEP} {\bfseries 10}
  (2013) 007}, \href{https://arxiv.org/abs/1210.7799}{{\ttfamily
  arXiv:1210.7799 [hep-th]}}.

\bibitem{Garcia-Etxebarria:2013tba}
I.~Garc\'\i{}a-Etxebarria, B.~Heidenreich, and T.~Wrase, ``{New N=1 dualities
  from orientifold transitions - Part II: String Theory},''
  \href{https://dx.doi.org/10.1007/JHEP10(2013)006}{{\em JHEP} {\bfseries 10}
  (2013) 006}, \href{https://arxiv.org/abs/1307.1701}{{\ttfamily
  arXiv:1307.1701 [hep-th]}}.

\bibitem{Etxebarria:2021lmq}
I.~Garc\'\i{}a-Etxebarria, B.~Heidenreich, M.~Lotito, and A.~K. Sorout,
  ``{Deconfining $ \mathcal{N} $ = 2 SCFTs or the art of brane bending},''
  \href{https://dx.doi.org/10.1007/JHEP03(2022)140}{{\em JHEP} {\bfseries 03}
  (2022) 140}, \href{https://arxiv.org/abs/2111.08022}{{\ttfamily
  arXiv:2111.08022 [hep-th]}}.

\bibitem{Bajeot:2022lah}
S.~Bajeot and S.~Benvenuti, ``{Sequential deconfinement and self-dualities in
  4d$ \mathcal{N} $ = 1 gauge theories},''
  \href{https://dx.doi.org/10.1007/JHEP10(2022)007}{{\em JHEP} {\bfseries 10}
  (2022) 007}, \href{https://arxiv.org/abs/2206.11364}{{\ttfamily
  arXiv:2206.11364 [hep-th]}}.

\bibitem{Bajeot:2022kwt}
S.~Bajeot and S.~Benvenuti, ``{S-confinements from deconfinements},''
  \href{https://dx.doi.org/10.1007/JHEP11(2022)071}{{\em JHEP} {\bfseries 11}
  (2022) 071}, \href{https://arxiv.org/abs/2201.11049}{{\ttfamily
  arXiv:2201.11049 [hep-th]}}.

\bibitem{Bottini:2022vpy}
L.~E. Bottini, C.~Hwang, S.~Pasquetti, and M.~Sacchi, ``{Dualities from
  dualities: the sequential deconfinement technique},''
  \href{https://dx.doi.org/10.1007/JHEP05(2022)069}{{\em JHEP} {\bfseries 05}
  (2022) 069}, \href{https://arxiv.org/abs/2201.11090}{{\ttfamily
  arXiv:2201.11090 [hep-th]}}.

\bibitem{Amariti:2023wts}
A.~Amariti, F.~Mantegazza, and D.~Morgante, ``{Sporadic dualities from tensor
  deconfinement},'' \href{https://arxiv.org/abs/2307.14146}{{\ttfamily
  arXiv:2307.14146 [hep-th]}}.

\bibitem{Amariti:2015kha}
A.~Amariti, C.~Cs\'aki, M.~Martone, and N.~R.-L. Lorier, ``{From 4D to 3D
  chiral theories: Dressing the monopoles},''
  \href{https://dx.doi.org/10.1103/PhysRevD.93.105027}{{\em Phys. Rev. D}
  {\bfseries 93} no.~10, (2016) 105027},
  \href{https://arxiv.org/abs/1506.01017}{{\ttfamily arXiv:1506.01017
  [hep-th]}}.

\bibitem{Nii:2016jzi}
K.~Nii, ``{3d Deconfinement, Product gauge group, Seiberg-Witten and New 3d
  dualities},'' \href{https://dx.doi.org/10.1007/JHEP08(2016)123}{{\em JHEP}
  {\bfseries 08} (2016) 123},
  \href{https://arxiv.org/abs/1603.08550}{{\ttfamily arXiv:1603.08550
  [hep-th]}}.

\bibitem{Pasquetti:2019uop}
S.~Pasquetti and M.~Sacchi, ``{From 3$d$ dualities to 2$d$ free field
  correlators and back},''
  \href{https://dx.doi.org/10.1007/JHEP11(2019)081}{{\em JHEP} {\bfseries 11}
  (2019) 081}, \href{https://arxiv.org/abs/1903.10817}{{\ttfamily
  arXiv:1903.10817 [hep-th]}}.

\bibitem{Pasquetti:2019tix}
S.~Pasquetti and M.~Sacchi, ``{3d dualities from 2d free field correlators:
  recombination and rank stabilization},''
  \href{https://dx.doi.org/10.1007/JHEP01(2020)061}{{\em JHEP} {\bfseries 01}
  (2020) 061}, \href{https://arxiv.org/abs/1905.05807}{{\ttfamily
  arXiv:1905.05807 [hep-th]}}.

\bibitem{Nii:2019dwi}
K.~Nii, ``{Confinement in 3d $\mathcal{N}=2$ exceptional gauge theories},''
  \href{https://arxiv.org/abs/1906.10161}{{\ttfamily arXiv:1906.10161
  [hep-th]}}.

\bibitem{Benvenuti:2020gvy}
S.~Benvenuti, I.~Garozzo, and G.~Lo~Monaco, ``{Sequential deconfinement in 3d $
  \mathcal{N} $ = 2 gauge theories},''
  \href{https://dx.doi.org/10.1007/JHEP07(2021)191}{{\em JHEP} {\bfseries 07}
  (2021) 191}, \href{https://arxiv.org/abs/2012.09773}{{\ttfamily
  arXiv:2012.09773 [hep-th]}}.

\bibitem{Benvenuti:2021nwt}
S.~Benvenuti and G.~Lo~Monaco, ``{A toolkit for ortho-symplectic dualities},''
  \href{https://arxiv.org/abs/2112.12154}{{\ttfamily arXiv:2112.12154
  [hep-th]}}.

\bibitem{Amariti:2022wae}
A.~Amariti and S.~Rota, ``{3d N=2 SO/USp adjoint SQCD: s-confinement and exact
  identities},'' \href{https://dx.doi.org/10.1016/j.nuclphysb.2022.116068}{{\em
  Nucl. Phys. B} {\bfseries 987} (2023) 116068},
  \href{https://arxiv.org/abs/2202.06885}{{\ttfamily arXiv:2202.06885
  [hep-th]}}.

\bibitem{Okazaki:2023hiv}
T.~Okazaki and D.~J. Smith, ``{Boundary confining dualities and Askey-Wilson
  type q-beta integrals},''
  \href{https://dx.doi.org/10.1007/JHEP08(2023)048}{{\em JHEP} {\bfseries 08}
  (2023) 048}, \href{https://arxiv.org/abs/2305.00247}{{\ttfamily
  arXiv:2305.00247 [hep-th]}}.

\bibitem{Sacchi:2020pet}
M.~Sacchi, ``{New 2d $ \mathcal{N} $ = (0, 2) dualities from four
  dimensions},'' \href{https://dx.doi.org/10.1007/JHEP12(2020)009}{{\em JHEP}
  {\bfseries 12} (2020) 009},
  \href{https://arxiv.org/abs/2004.13672}{{\ttfamily arXiv:2004.13672
  [hep-th]}}.

\bibitem{Gadde:2013wq}
A.~Gadde, S.~Gukov, and P.~Putrov, ``{Walls, Lines, and Spectral Dualities in
  3d Gauge Theories},'' \href{https://dx.doi.org/10.1007/JHEP05(2014)047}{{\em
  JHEP} {\bfseries 1405} (2014) 047},
\href{https://arxiv.org/abs/1302.0015}{{\ttfamily arXiv:1302.0015 [hep-th]}}.
%%CITATION = ARXIV:1302.0015;%%.

\bibitem{Okazaki:2013kaa}
T.~Okazaki and S.~Yamaguchi, ``{Supersymmetric boundary conditions in
  three-dimensional N=2 theories},''
  \href{https://dx.doi.org/10.1103/PhysRevD.87.125005}{{\em Phys.Rev.}
  {\bfseries D87} no.~12, (2013) 125005},
\href{https://arxiv.org/abs/1302.6593}{{\ttfamily arXiv:1302.6593 [hep-th]}}.
%%CITATION = ARXIV:1302.6593;%%.

\bibitem{Gadde:2013sca}
A.~Gadde, S.~Gukov, and P.~Putrov, ``{Fivebranes and 4-manifolds},''
\href{https://arxiv.org/abs/1306.4320}{{\ttfamily arXiv:1306.4320 [hep-th]}}.
%%CITATION = ARXIV:1306.4320;%%.

\bibitem{Yoshida:2014ssa}
Y.~Yoshida and K.~Sugiyama, ``{Localization of three-dimensional
  $\mathcal{N}=2$ supersymmetric theories on $S^1 \times D^2$},''
  \href{https://dx.doi.org/10.1093/ptep/ptaa136}{{\em PTEP} {\bfseries 2020}
  no.~11, (2020) 113B02}, \href{https://arxiv.org/abs/1409.6713}{{\ttfamily
  arXiv:1409.6713 [hep-th]}}.

\bibitem{Dimofte:2017tpi}
T.~Dimofte, D.~Gaiotto, and N.~M. Paquette, ``{Dual boundary conditions in 3d
  SCFT's},'' \href{https://dx.doi.org/10.1007/JHEP05(2018)060}{{\em JHEP}
  {\bfseries 05} (2018) 060},
\href{https://arxiv.org/abs/1712.07654}{{\ttfamily arXiv:1712.07654 [hep-th]}}.
%%CITATION = ARXIV:1712.07654;%%.

\bibitem{Brunner:2019qyf}
I.~Brunner, J.~Schulz, and A.~Tabler, ``{Boundaries and supercurrent multiplets
  in 3D Landau-Ginzburg models},''
  \href{https://dx.doi.org/10.1007/JHEP06(2019)046}{{\em JHEP} {\bfseries 06}
  (2019) 046}, \href{https://arxiv.org/abs/1904.07258}{{\ttfamily
  arXiv:1904.07258 [hep-th]}}.

\bibitem{Costello:2020ndc}
K.~Costello, T.~Dimofte, and D.~Gaiotto, ``{Boundary Chiral Algebras and
  Holomorphic Twists},'' \href{https://arxiv.org/abs/2005.00083}{{\ttfamily
  arXiv:2005.00083 [hep-th]}}.

\bibitem{Sugiyama:2020uqh}
K.~Sugiyama and Y.~Yoshida, ``{Supersymmetric indices on $I \times T^2$,
  elliptic genera and dualities with boundaries},''
  \href{https://dx.doi.org/10.1016/j.nuclphysb.2020.115168}{{\em Nucl. Phys. B}
  {\bfseries 960} (2020) 115168},
  \href{https://arxiv.org/abs/2007.07664}{{\ttfamily arXiv:2007.07664
  [hep-th]}}.

\bibitem{Okazaki:2021pnc}
T.~Okazaki and D.~J. Smith, ``{Seiberg-like dualities for orthogonal and
  symplectic 3d $ \mathcal{N} $ = 2 gauge theories with boundaries},''
  \href{https://dx.doi.org/10.1007/JHEP07(2021)231}{{\em JHEP} {\bfseries 07}
  (2021) 231}, \href{https://arxiv.org/abs/2105.07450}{{\ttfamily
  arXiv:2105.07450 [hep-th]}}.

\bibitem{Okazaki:2021gkk}
T.~Okazaki and D.~J. Smith, ``{Web of Seiberg-like dualities for 3D N=2
  quivers},'' \href{https://dx.doi.org/10.1103/PhysRevD.105.086023}{{\em Phys.
  Rev. D} {\bfseries 105} no.~8, (2022) 086023},
  \href{https://arxiv.org/abs/2112.07347}{{\ttfamily arXiv:2112.07347
  [hep-th]}}.

\bibitem{Alekseev:2022gnr}
S.~Alekseev, M.~Dedushenko, and M.~Litvinov, ``{Chiral life on a slab},''
  \href{https://arxiv.org/abs/2301.00038}{{\ttfamily arXiv:2301.00038
  [hep-th]}}.

\bibitem{Dedushenko:2022fmc}
M.~Dedushenko and M.~Litvinov, ``{Interval reduction and (super)symmetry},''
  \href{https://arxiv.org/abs/2212.07455}{{\ttfamily arXiv:2212.07455
  [hep-th]}}.

\bibitem{Crew:2023tky}
S.~Crew, D.~Zhang, and B.~Zhao, ``{Boundaries \& Localisation with a
  Topological Twist},'' \href{https://arxiv.org/abs/2306.16448}{{\ttfamily
  arXiv:2306.16448 [hep-th]}}.

\bibitem{Dedushenko:2023qjq}
M.~Dedushenko and N.~Nekrasov, ``{Interfaces and Quantum Algebras, II: Cigar
  Partition Function},'' \href{https://arxiv.org/abs/2306.16434}{{\ttfamily
  arXiv:2306.16434 [hep-th]}}.

\bibitem{MR783216}
R.~Askey and J.~Wilson, ``Some basic hypergeometric orthogonal polynomials that
  generalize {J}acobi polynomials,''
  \href{https://dx.doi.org/10.1090/memo/0319}{{\em Mem. Amer. Math. Soc.}
  {\bfseries 54} no.~319, (1985) iv+55}.
  \url{https://doi.org/10.1090/memo/0319}.

\bibitem{MR772878}
B.~Nassrallah and M.~Rahman, ``Projection formulas, a reproducing kernel and a
  generating function for {$q$}-{W}ilson polynomials,''
  \href{https://dx.doi.org/10.1137/0516014}{{\em SIAM J. Math. Anal.}
  {\bfseries 16} no.~1, (1985) 186--197}.
  \url{https://doi.org/10.1137/0516014}.

\bibitem{MR845667}
M.~Rahman, ``An integral representation of a {$_{10}\varphi_9$} and continuous
  bi-orthogonal {$_{10}\varphi_9$} rational functions,''
  \href{https://dx.doi.org/10.4153/CJM-1986-030-6}{{\em Canad. J. Math.}
  {\bfseries 38} no.~3, (1986) 605--618}.
  \url{https://doi.org/10.4153/CJM-1986-030-6}.

\bibitem{MR1139492}
R.~A. Gustafson, ``Some {$q$}-beta and {M}ellin-{B}arnes integrals on compact
  {L}ie groups and {L}ie algebras,''
  \href{https://dx.doi.org/10.2307/2154615}{{\em Trans. Amer. Math. Soc.}
  {\bfseries 341} no.~1, (1994) 69--119}.
  \url{https://doi.org/10.2307/2154615}.

\bibitem{MR1266569}
R.~A. Gustafson, ``Some {$q$}-beta integrals on {${\rm SU}(n)$} and {${\rm
  Sp}(n)$} that generalize the {A}skey-{W}ilson and {N}asrallah-{R}ahman
  integrals,'' \href{https://dx.doi.org/10.1137/S0036141092248614}{{\em SIAM J.
  Math. Anal.} {\bfseries 25} no.~2, (1994) 441--449}.
  \url{https://doi.org/10.1137/S0036141092248614}.

\bibitem{MR2267266}
M.~Ito, ``Askey-{W}ilson type integrals associated with root systems,''
  \href{https://dx.doi.org/10.1007/s11139-006-9579-y}{{\em Ramanujan J.}
  {\bfseries 12} no.~1, (2006) 131--151}.
  \url{https://doi.org/10.1007/s11139-006-9579-y}.

\bibitem{Ramond:1996ku}
P.~Ramond, ``{Superalgebras in N=1 gauge theories},''
  \href{https://dx.doi.org/10.1016/S0370-2693(96)01420-7}{{\em Phys. Lett. B}
  {\bfseries 390} (1997) 179--184},
  \href{https://arxiv.org/abs/hep-th/9608077}{{\ttfamily
  arXiv:hep-th/9608077}}.

\bibitem{Distler:1996ub}
J.~Distler and A.~Karch, ``{N=1 dualities for exceptional gauge groups and
  quantum global symmetries},''
  \href{https://dx.doi.org/10.1002/prop.2190450603}{{\em Fortsch. Phys.}
  {\bfseries 45} (1997) 517--533},
  \href{https://arxiv.org/abs/hep-th/9611088}{{\ttfamily
  arXiv:hep-th/9611088}}.

\bibitem{Karch:1997jp}
A.~Karch, ``{More on N=1 selfdualities and exceptional gauge groups},''
  \href{https://dx.doi.org/10.1016/S0370-2693(97)00604-7}{{\em Phys. Lett. B}
  {\bfseries 405} (1997) 280--286},
  \href{https://arxiv.org/abs/hep-th/9702179}{{\ttfamily
  arXiv:hep-th/9702179}}.

\bibitem{Cho:1997am}
P.~L. Cho, ``{Moduli in exceptional SUSY gauge theories},''
  \href{https://dx.doi.org/10.1103/PhysRevD.57.5214}{{\em Phys. Rev. D}
  {\bfseries 57} (1998) 5214--5223},
  \href{https://arxiv.org/abs/hep-th/9712116}{{\ttfamily
  arXiv:hep-th/9712116}}.

\bibitem{Grinstein:1998bu}
B.~Grinstein and D.~R. Nolte, ``{Systematic study of theories with quantum
  modified moduli. 2.},''
  \href{https://dx.doi.org/10.1103/PhysRevD.58.045012}{{\em Phys. Rev. D}
  {\bfseries 58} (1998) 045012},
  \href{https://arxiv.org/abs/hep-th/9803139}{{\ttfamily
  arXiv:hep-th/9803139}}.

\bibitem{Pouliot:2001iw}
P.~Pouliot, ``{Spectroscopy of gauge theories based on exceptional Lie
  groups},'' \href{https://dx.doi.org/10.1088/0305-4470/34/41/317}{{\em J.
  Phys. A} {\bfseries 34} (2001) 8631--8658},
  \href{https://arxiv.org/abs/hep-th/0107151}{{\ttfamily
  arXiv:hep-th/0107151}}.

\bibitem{Nii:2017npz}
K.~Nii and Y.~Sekiguchi, ``{Low-energy dynamics of 3d $ \mathcal{N} $ = 2
  G$_{2}$ supersymmetric gauge theory},''
  \href{https://dx.doi.org/10.1007/JHEP02(2018)158}{{\em JHEP} {\bfseries 02}
  (2018) 158}, \href{https://arxiv.org/abs/1712.02774}{{\ttfamily
  arXiv:1712.02774 [hep-th]}}.

\bibitem{Nii:2019wjz}
K.~Nii, ``{\textquotedblleft{}Chiral\textquotedblright{} and
  \textquotedblleft{}non-chiral\textquotedblright{} 3d Seiberg duality},''
  \href{https://dx.doi.org/10.1007/JHEP04(2020)098}{{\em JHEP} {\bfseries 04}
  (2020) 098}, \href{https://arxiv.org/abs/1907.03340}{{\ttfamily
  arXiv:1907.03340 [hep-th]}}.

\bibitem{Chen:2018wep}
Z.~Chen, W.~Gu, H.~Parsian, and E.~Sharpe, ``{Two-dimensional supersymmetric
  gauge theories with exceptional gauge groups},''
  \href{https://dx.doi.org/10.4310/ATMP.2020.v24.n1.a3}{{\em Adv. Theor. Math.
  Phys.} {\bfseries 24} no.~1, (2020) 67--123},
  \href{https://arxiv.org/abs/1808.04070}{{\ttfamily arXiv:1808.04070
  [hep-th]}}.

\bibitem{MR674768}
I.~G. Macdonald, ``Some conjectures for root systems,''
  \href{https://dx.doi.org/10.1137/0513070}{{\em SIAM J. Math. Anal.}
  {\bfseries 13} no.~6, (1982) 988--1007}.
  \url{https://doi.org/10.1137/0513070}.

\bibitem{MR1314036}
I.~Cherednik, ``Double affine {H}ecke algebras and {M}acdonald's conjectures,''
  \href{https://dx.doi.org/10.2307/2118632}{{\em Ann. of Math. (2)} {\bfseries
  141} no.~1, (1995) 191--216}. \url{https://doi.org/10.2307/2118632}.

\bibitem{MR2018362}
I.~G. Macdonald, \href{https://dx.doi.org/10.1090/trans2/210/14}{``A formal
  identity for affine root systems,''} in {\em Lie groups and symmetric
  spaces}, vol.~210 of {\em Amer. Math. Soc. Transl. Ser. 2}, pp.~195--211.
\newblock Amer. Math. Soc., Providence, RI, 2003.
\newblock \url{https://doi.org/10.1090/trans2/210/14}.

\bibitem{MR1868358}
M.~Ito, ``Symmetry classification for {J}ackson integrals associated with
  irreducible reduced root systems,''
  \href{https://dx.doi.org/10.1023/A:1012518910847}{{\em Compositio Math.}
  {\bfseries 129} no.~3, (2001) 325--340}.
  \url{https://doi.org/10.1023/A:1012518910847}.

\bibitem{MR1926355}
M.~Ito, ``A product formula for {J}ackson integral associated with the root
  system {$F_4$},'' \href{https://dx.doi.org/10.1023/A:1019711531212}{{\em
  Ramanujan J.} {\bfseries 6} no.~3, (2002) 279--293}.
  \url{https://doi.org/10.1023/A:1019711531212}.

\bibitem{MR2016669}
M.~Ito, ``Convergence and asymptotic behavior of {J}ackson integrals associated
  with irreducible reduced root systems,''
  \href{https://dx.doi.org/10.1016/j.jat.2003.08.006}{{\em J. Approx. Theory}
  {\bfseries 124} no.~2, (2003) 154--180}.
  \url{https://doi.org/10.1016/j.jat.2003.08.006}.

\bibitem{Cordova:2018qvg}
C.~C\'ordova, P.-S. Hsin, and K.~Ohmori, ``{Exceptional Chern-Simons-Matter
  Dualities},'' \href{https://dx.doi.org/10.21468/SciPostPhys.7.4.056}{{\em
  SciPost Phys.} {\bfseries 7} no.~4, (2019) 056},
  \href{https://arxiv.org/abs/1812.11705}{{\ttfamily arXiv:1812.11705
  [hep-th]}}.

\bibitem{Okazaki:2019bok}
T.~Okazaki, ``{Abelian dualities of $\mathcal{N}=(0,4)$ boundary conditions},''
  \href{https://dx.doi.org/10.1007/JHEP08(2019)170}{{\em JHEP} {\bfseries 08}
  (2019) 170},
\href{https://arxiv.org/abs/1905.07425}{{\ttfamily arXiv:1905.07425 [hep-th]}}.
%%CITATION = ARXIV:1905.07425;%%.

\bibitem{Gaiotto:2008ak}
D.~Gaiotto and E.~Witten, ``{S-Duality of Boundary Conditions In N=4 Super
  Yang-Mills Theory},''
  \href{https://dx.doi.org/10.4310/ATMP.2009.v13.n3.a5}{{\em Adv. Theor. Math.
  Phys.} {\bfseries 13} no.~3, (2009) 721--896},
\href{https://arxiv.org/abs/0807.3720}{{\ttfamily arXiv:0807.3720 [hep-th]}}.
%%CITATION = ARXIV:0807.3720;%%.

\bibitem{Okazaki:2020lfy}
T.~Okazaki, ``{Abelian mirror symmetry of $ \mathcal{N} $ = (2, 2) boundary
  conditions},'' \href{https://dx.doi.org/10.1007/JHEP03(2021)163}{{\em JHEP}
  {\bfseries 03} (2021) 163},
  \href{https://arxiv.org/abs/2010.13177}{{\ttfamily arXiv:2010.13177
  [hep-th]}}.

\bibitem{Gaiotto:2019jvo}
D.~Gaiotto and T.~Okazaki, ``{Dualities of Corner Configurations and
  Supersymmetric Indices},''
  \href{https://dx.doi.org/10.1007/JHEP11(2019)056}{{\em JHEP} {\bfseries 11}
  (2019) 056},
\href{https://arxiv.org/abs/1902.05175}{{\ttfamily arXiv:1902.05175 [hep-th]}}.
%%CITATION = ARXIV:1902.05175;%%.

\bibitem{Pesando:1995bq}
I.~Pesando, ``{Exact results for the supersymmetric G(2) gauge theories},''
  \href{https://dx.doi.org/10.1142/S0217732395002027}{{\em Mod. Phys. Lett. A}
  {\bfseries 10} (1995) 1871--1886},
  \href{https://arxiv.org/abs/hep-th/9506139}{{\ttfamily
  arXiv:hep-th/9506139}}.

\bibitem{Giddings:1995ns}
S.~B. Giddings and J.~M. Pierre, ``{Some exact results in supersymmetric
  theories based on exceptional groups},''
  \href{https://dx.doi.org/10.1103/PhysRevD.52.6065}{{\em Phys. Rev. D}
  {\bfseries 52} (1995) 6065--6073},
  \href{https://arxiv.org/abs/hep-th/9506196}{{\ttfamily
  arXiv:hep-th/9506196}}.

\end{thebibliography}\endgroup

\end{document}